\pdfoutput=1	

\documentclass[10pt,twocolumn]{article}

\usepackage[left=0.88in, right=0.88in, top=1in, bottom=1in, columnsep=0.25in]{geometry}
\usepackage{amsmath}
\usepackage{bm}	
\usepackage{graphicx}
\usepackage{hyperref}

\title{\emph{Young TIM}: A wave-optics simulator with slightly special powers}

\author{Sean Leavey 	
and Johannes Courtial}

\begin{document}

\maketitle

\begin{abstract}
Wave optics is a prominent part of the undergraduate physics curriculum, and many undergraduate labs contain experiments on wave optics.
In our 3rd-year undergraduate lab, we run numerical simulations alongside the experiments (and when pandemic restrictions did not allow students into the lab, those simulations  replaced some experiments).
We use \emph{Young TIM}, an interactive wave-optics simulator designed to be a research tool that can also be used for the dissemination of our research and for education.
It has novel and unique features, including the ability to create an anaglyph of the beam as it would be seen in 3D by a binocular, probably misguided, observer staring into the beam.
Here we describe how to use \emph{Young TIM}, and we describe several possible numerical experiments suitable for undergraduate teaching.
\end{abstract}



\newcommand{\rmi}{\mathrm{i}}
\newcommand{\rmd}{\mathrm{d}}
\newcommand{\bi}[1]{\mathbf{#1}}
\newcommand{\p}{^\prime}
\newcommand{\TODO}[1]{\textcolor{red}{[TODO: #1]}}

\renewcommand\floatpagefraction{.999}
\renewcommand\topfraction{.999}
\renewcommand\bottomfraction{.999}
\renewcommand\textfraction{.001}


\section{Introduction}

\noindent
Wave optics is an important part of the undergraduate physics curriculum.
Not only does it teach the basics of optics, but it also gives students experience with waves, which are ubiquitous in physics, and it is the prime example of a description of a phenomenon giving a ``deeper'' understanding of another description of the same phenomenon, in this case ray optics.
Understanding wave optics is important for understanding vital concepts such as the resolution of optical telescopes and microscopes.
Wave optics is, at the same time, a very old field of science (the concept of light being a wave is included in Huygens' \emph{Trait\'{e} de la Lumi\`{e}re} \cite{Huygens-1690}, published in 1690) and an active area of research, recently receiving a boost with the advent of metamaterials \cite{Zheludev-Kivshar-2012}, man-made materials with bespoke optical (including wave-optical) properties, and the invention of transformation optics \cite{Ward-Pendry-1996,Pendry-et-al-2006,Leonhardt-2006}, a powerful way of harnessing the new possibilities unlocked by metamaterials.


It is also relatively easy and cheap to perform wave-optics experiments, and for this reason many undergraduate laboratories, including ours, contain experiments on wave optics.
A number of years ago, we started to use \emph{Young TIM} in our 3rd-year undergraduate physics laboratory to accompany experiments on Fraunhofer diffraction and laser modes.
This enables observations that are otherwise difficult or cumbersome to perform in an undergraduate lab, ranging from a visualisation of the transverse phase cross-section of a light beam (which would require interferometry in an experiment) and longitudinal beam cross sections (which would require scanning of the beam in a large number of transverse planes) to the evolution of a light beam over successive round trips through a laser resonator into the lowest-loss eigenmode (which requires heroic efforts even in a research laboratory).
In our undergraduate lab, where students work in teams, running simulations alongside experiments also means that students can work on numerical tasks at those times when other members of their team work on the experiment.
When in-person experiments were not allowed in our undergraduate laboratories due to pandemic restrictions, the numerical simulations largely replaced the experiments.

For our numerical simulations, we use \emph{Young TIM}.
We designed \emph{Young TIM} to be at the same time a specialised research tool and an easy-to-use tool for teaching wave optics.
In contrast to other free wave-optics simulation tools used for teaching, such as PhET \cite{PhET-wave-interference} and virtual lab simulations such as those described in Refs \cite{Fernandez-et-al-2014,Baranov-2018}, \emph{Young TIM} is less of a simulator for exploring specific aspects of wave optics, and more of a well-equipped virtual laboratory, representing a free alternative to the use of commercial software such as Comsol in teaching \cite{Yang-et-al-2008a}.
Here we describe how to use \emph{Young TIM}, and we describe several suggested numerical experiments, specifically double-slit diffraction, laser-mode formation, and holography.

\emph{Young TIM}'s basic functionality is to represent a light beam's field in a transverse plane, and to simulate the change in this field on transmission through an optical system, which consists of a number of optical components.
\emph{Young TIM} is currently restricted to scalar waves, i.e.\ it ignores any effect related to polarisation.
The (scalar) light field at a point is represented by its complex amplitude, a complex number whose modulus represents the field amplitude and whose argument represents the field's phase.
The light field in a transverse plane is represented in the form of the values of the complex field on a Cartesian grid of points that cover a rectangular area in that plane.
\emph{Young TIM} can simulate transmission through various standard optical components, including thin lenses \cite{Goodman-1996-lens-phase}, beam splitters, aperture stops, and thin holograms.
Propagation through the (linear) medium between the components is calculated using a (non-paraxial) version of a well-known Fourier algorithm~\cite{Sziklas-Siegman-1975}.


As \emph{Young TIM} is also designed to be a tool to support various aspects of our research, it is also able to handle specialised light beams and optical components we use in this research.
Examples include 
Laguerre-Gaussian beams and Dove prisms, which we use in our research on optical angular momentum \cite{Leach-et-al-2002}, and Dove-prism arrays \cite{Hamilton-Courtial-2008a} and lenslet (or microlens) arrays, which we use in our research on pixellated optical components \cite{Hamilton-Courtial-2009,Courtial-2008a,Bourgenot-et-al-2018}.
As an aside, the wave optics of pixellated optical components is of particular interest to us as these components are designed such that they \emph{appear to} --- but do not actually --- change the direction of transmitted light rays in ways not allowed by wave optics
\cite{Hamilton-Courtial-2009}.

A number of \emph{Young TIM}'s features make it useful as a teaching tool.
Firstly, it is designed to be interactive and easy to use.
It is also designed to be fun to use, for example by being able to create anaglyphs that visualise the 3D experience of looking into the beam with both eyes.
Secondly, it is written in Java, which means it can be run on any computer running a reasonably up-to-date installation of Java.
Finally, \emph{Young TIM} is open-source software --- the source code can be freely downloaded \cite{YoungTIM-source} 
--- , and so particularly keen students can inspect, alter or extend \emph{Young TIM}'s source code, for example in order to add their own optical components.

The level of interactivity in \emph{Young TIM} can be varied.
Specifically, \emph{Young TIM} has been developed such that its code can be compiled into different versions:
\begin{enumerate}
\item a non-interactive (command-line) version, mainly for research purposes, but which also means that \emph{Young TIM}'s use in teaching wave optics could include a coding element, similar to the approach taken in Ref.\ \cite{Yim-Lee-1993};
\item a ``slightly interactive'' version, which limits interactive control to specific parameters, and which can be used to illustrate specific aspects of wave optics;
\item a fully interactive version that invites playful exploration of wave optics.
\end{enumerate}
Unless otherwise stated, we discuss here the fully interactive version.

We have previously written ray-tracing software called TIM which allows us to simulate somewhat ``photorealistic'' images~\cite{Lambert-et-al-2012,Oxburgh-et-al-2014}, specifically of the view through microstructured optical components of interest in our research\footnote{Those microstructured optical components are sometimes called \underline{meta}ma\underline{t}erials f\underline{o}r light ra\underline{y}s (METATOYs) \cite{Hamilton-Courtial-2009}, and TIM is an acronym for \underline{T}he \underline{I}nteractive \underline{M}ETATOY.}.
The software we describe here can be seen as the wave-optics analogue of TIM.
As wave optics is perhaps best illustrated by Thomas Young's double-slit experiment, and as we developed the software we describe here after we developed TIM (and it is therefore younger), we call our new software \emph{Young TIM}.

This paper is structured as follows.
Section \ref{s:usingYoungTIM} describes how to use \emph{Young TIM}.
In section \ref{exercises-section}, we describe a few suggested numerical experiments students can perform using \emph{Young TIM}.
One of \emph{Young TIM}'s special powers, namely the ability to create anaglyphs (3D images for use with red/blue glasses) from wave fields, is described in section \ref{anaglyphs-section}.
We discuss some of the common numerical artefacts that occur when simulating free-space propagation of a light wave in section \ref{artefacts-section}), before concluding.
Appendix \ref{components-section} is a reference guide to \emph{Young TIM}'s optical components.
The other appendices give a few details intended for users of the non-interactive --- and more versatile --- version of \emph{Young TIM}.
Appendix \ref{source-code-modification-appendix} is a brief introduction into a few of the more common source-code modification tasks, and appendix \ref{non-planar-surfaces-appendix} is a brief outline of \emph{Young TIM}'s capability to simulate propagation between non-planar surfaces, which is not available in the interactive version.

\section{Using \emph{Young TIM}\label{s:usingYoungTIM}}

\subsection{Simulating the default optical system\label{default-optical-system-section}}

\begin{figure}
\begin{center} \includegraphics[width=\columnwidth]{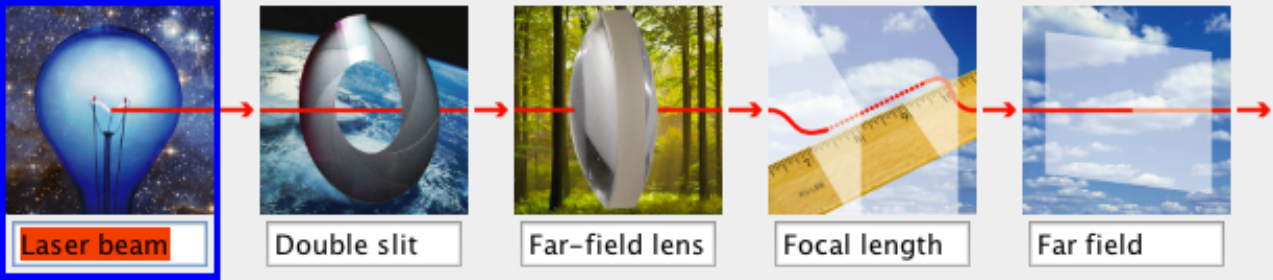} \end{center}
\caption{\label{default-optical-environment-figure}\emph{Young TIM}'s graphical representation of its default startup optical system.
The optical system consists of a light beam, a double slit, a lens, a component representing the distance between the lens and its image-sided focal plane, and a plane component that marks this image-sided focal plane.
The plane component allows the beam cross-section in that plane to be inspected;
as the lens produces an image of the double slit's far field in its image-sided focal plane, the plane is labelled ``Far field''.
This optical system allows experimentation with double-slit diffraction.
All optical elements in \emph{Young TIM} have a user-defined name, allowing labelling that is meaningful for the user.
The blue frame around the ``Laser beam'' component means that this component's parameters are shown in the panel underneath (not shown here; see Fig.\ \ref{light-source-panel-figure} for an example).
There is no special meaning in the choice of background images, which were chosen to appeal to one of the authors' four-year old daughter.
}
\end{figure}

\noindent
Fig.\ \ref{default-optical-environment-figure} shows the graphical representation of the optical system contained in the fully interactive version of \emph{Young TIM} on startup, namely Young's double-slit experiment.
This is shown at the top of \emph{Young TIM}'s screen area.
The optical system consists of a number of elements:
a laser beam, a double slit, a lens, and two elements that represent the distance between the lens and its image-sided focal plane, and the image-sided focal plane.
Clicking on the ``Go!'' button at the bottom simulates light propagation through the optical system.
Clicking on any element shows a panel underneath that allows inspecting and editing that element's parameters or, in the case of a plane, visualising the simulated wave there.

\begin{figure}
\begin{center} \includegraphics[width=\columnwidth]{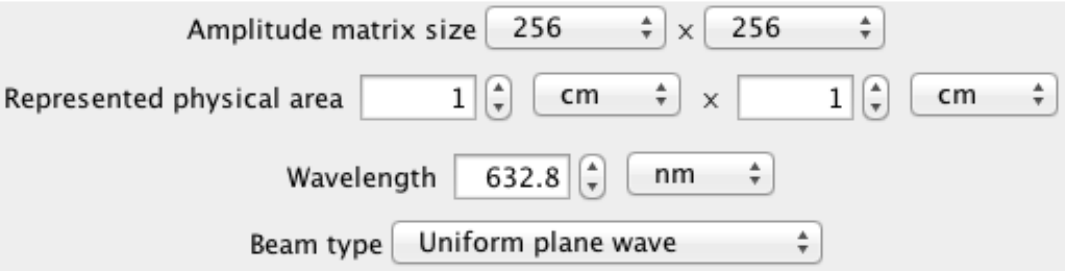} \end{center}
\caption{\label{light-source-panel-figure}Panel for editing the parameters of the initial beam.
The size of the amplitude matrix and of the represented physical area determine how the beam is represented internally (see Fig.\ \ref{grid-of-points-figure}).
The remaining parameters determine the beam's wavelength and type.}
\end{figure}

\begin{figure}
\begin{center} \includegraphics{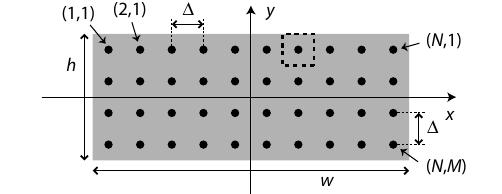} \end{center}
\caption{\label{grid-of-points-figure}Amplitude matrix and represented physical area.
The elements in the amplitude matrix represent samples of the beam's complex amplitude on a Cartesian grid of $N \times M$ points in the $(x, y)$ plane.
The spacing between neighbouring grid points in the $x$ and $y$ direction is the same, namely $\Delta$.
The array of black dots shown indicates such a grid of points; a few of these points are labelled with the indices $(i, j)$ of the corresponding matrix elements.
The dashed rectangle surrounds the area represented by one particular grid point.
The physical area represented by the entire amplitude matrix is shaded; its width and height is respectively $w$ and $h$.
The origin of the $(x, y)$ coordinate system is centred between the four central points.}
\end{figure}

Fig.\ \ref{light-source-panel-figure} shows an example of those parameter panels, namely the panel for editing the laser beam's parameters.
The amplitude matrix size and represented physical area describe the way the beam's complex amplitude is internally represented, i.e.\ the number of columns and rows in the amplitude matrix, and the width and height of the physical area represented by the amplitude matrix (see Fig.\ \ref{grid-of-points-figure}).
The other parameters specify the beam's wavelength, the type of beam, and the parameters for the selected beam type.

\begin{figure}
\begin{center} \includegraphics[width=\columnwidth]{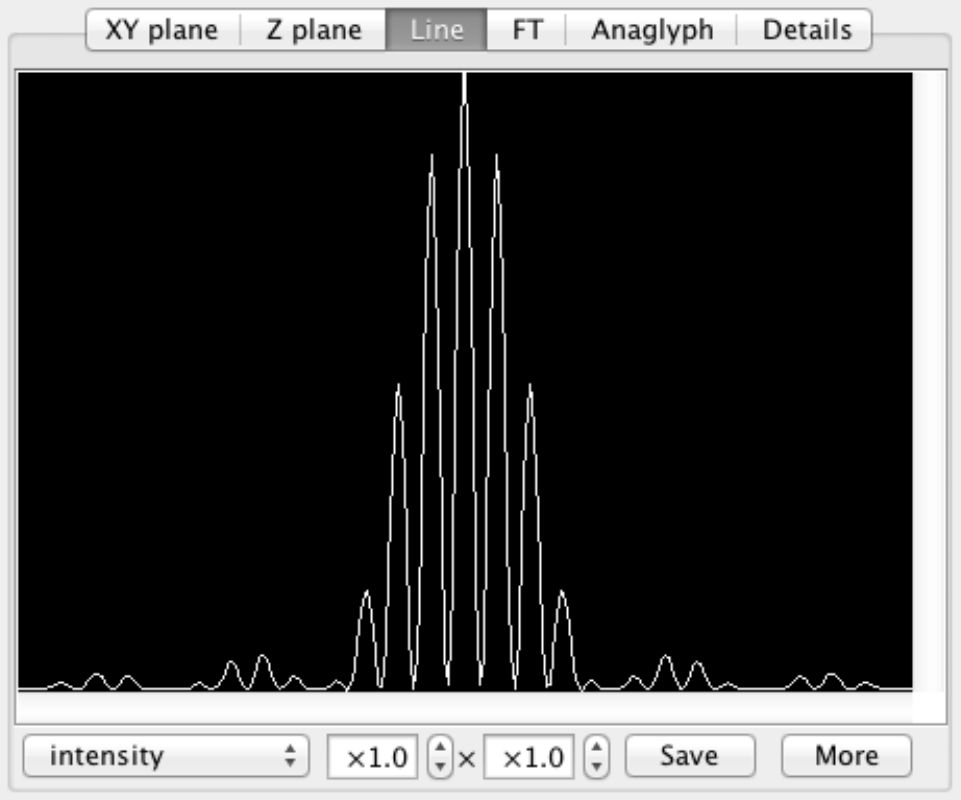} \end{center}
\caption{\label{plane-panel-figure}Panel of the plane representing the far field in the default startup optical system after propagation through the optical system has been simulated.
The main part of the panel is taken up by a tabs.
Selecting the ``Details'' tab shows a list of parameters describing the beam in the plane, including the power in the beam;
selecting any of the other tabs shows a graphical representation of the beam cross-section.
In the example shown here, an intensity cross-section along the $x$ axis (i.e.\ the line $y = 0$) is shown.
}
\end{figure}

\begin{figure}
\begin{center}
\includegraphics[width=\columnwidth]{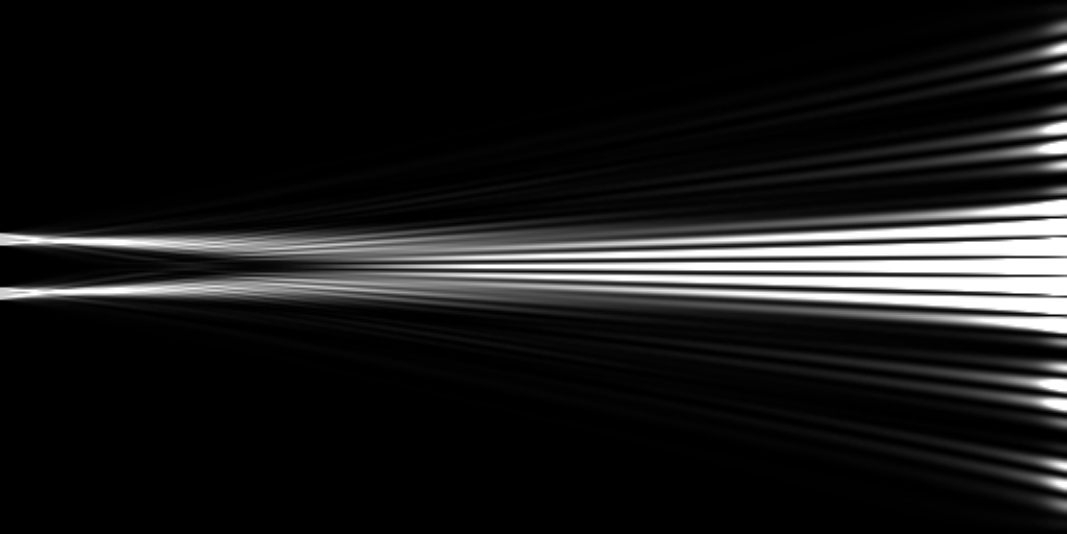}
\end{center}
\caption{\label{Z-plane-figure}Example of a plot of the beam in a longitudinal plane.
Here, the intensity of the beam in the default optical system is shown in the $(x, z)$ plane ($y = 0$) between the far-field lens and its focal plane, i.e.\ from $\Delta z_\mathrm{min} = -60\,\mathrm{cm}$ to $\Delta z_\mathrm{max} = 0$, where $\Delta z$ is measured relative to the position of the far-field lens's focal plane.
The exposure compensation is set to $+5.5$ to show up interesting features in the beam.
To create a plot such as this, select the \texttt{Z plane} tab, press the \texttt{More} button, and adjust the parameters.}
\end{figure}

The plane is a special element that does not in any way alter the beam passing through it.
In \emph{Young TIM}, a plane has two functions:
\begin{enumerate}
\item It allows the beam cross-section in that plane to be examined.
Fig.\ \ref{plane-panel-figure} shows the panel of the plane representing the far field in the default optical system.
The beam can also be examined \emph{around} that plane.
Fig.\ \ref{Z-plane-figure} shows the intensity of the beam in the $(x, z)$ plane ($z$ is the propagation direction) between the far-field lens behind the slits and its focal plane.
\item It can be imaged to another position in the optical train.
This allows the representation of optical systems with a non-linear topology, for example resonators and interferometers (see section \ref{other-optical-systems-section}).
\end{enumerate}

\subsection{Representing and simulating other optical systems\label{other-optical-systems-section}}

\noindent
\emph{Young TIM} can handle optical systems considerably more general than aperture diffraction as discussed in section \ref{default-optical-system-section}.
Here we discuss an example that illustrates a few additional capabilities.

\begin{figure}
\begin{center}
\includegraphics{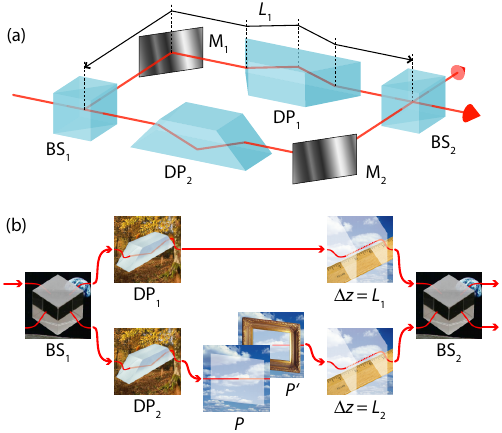}
\end{center}
\caption{\label{OAM-sorter-figure}Diagrammatic representation~(a) and representation in terms of \emph{Young TIM}'s components~(b) of an example of an optical setup that can be simulated in \emph{Young TIM}.
(a)~The Mach-Zehnder interferometer shown here, which has a Dove prism in each arm, can sort light into different orbital-angular-momentum (OAM) states \cite{Leach-et-al-2002,Wei-et-al-2003}.
(b)~Solid red arrows represent beams passing between components.
Non-zero distance between components is represented by a special \texttt{distance} optical component.
Two planes in the setup can be effectively linked with a pair of special optical components, one (of type \texttt{Plane}) that defines a plane, and another (of type \texttt{Image of plane}) that refers to a 1:1 image of the plane.
The lower arm of the interferometer contains such a link between a plane $P$ and its image $P^\prime$.}
\end{figure}

As our example we have chosen the optical setup shown in Fig.\ \ref{OAM-sorter-figure}(a).
It is a setup we have used in our own research \cite{Leach-et-al-2002}, so it serves also as an example of the potential of \emph{Young TIM} to be applied to research.
This setup contains a few optical components that are not in the startup optical system, but perhaps the most interesting difference is that, by virtue of being a two-arm interferometer, its topology is different from the simple linear topology of the startup optical system.

The optical system can be altered by removing optical elements and by inserting optical elements.
Components can be removed individually from the optical system by right-clicking on the component, which brings up a popup menu, and selecting ``Remove''.
(Note that there is also a possibility to disable individual components.
This functionality is intended to allow easy temporary removal of components from the optical system.)
To insert a component at a specific point in the optical system, right-click on the arrow at the point in the optical system where the component is to be inserted (between, in front of, or behind components), and select the component to be inserted from the popup menu that appears.

A general non-linear topology can be created in \emph{Young TIM} by first creating a tree topology, which simply allows light beams to be combined and/or to be split, and by then connecting different points in the tree.
The tree topology is created by inserting optical components that have more than one input and/or output, for example a beam splitter, which has two inputs and two outputs.
Connections between different points in the tree are made by inserting planes and images of those planes.
Fig.\ \ref{OAM-sorter-figure}(b) outlines how the interferometer shown in Fig.\ \ref{OAM-sorter-figure}(a) can be represented in this way.

Starting from an empty optical system, the interferometer shown in Fig.\ \ref{OAM-sorter-figure} can be created in \emph{Young TIM} as follows.
The components of the new optical system have to be added, one by one, through insertion at their place in the optical system.
We can start with any component in the optical system;
here we start with the first beam splitter, BS$_1$.

\begin{figure*}
\begin{center} \includegraphics[width=\textwidth]{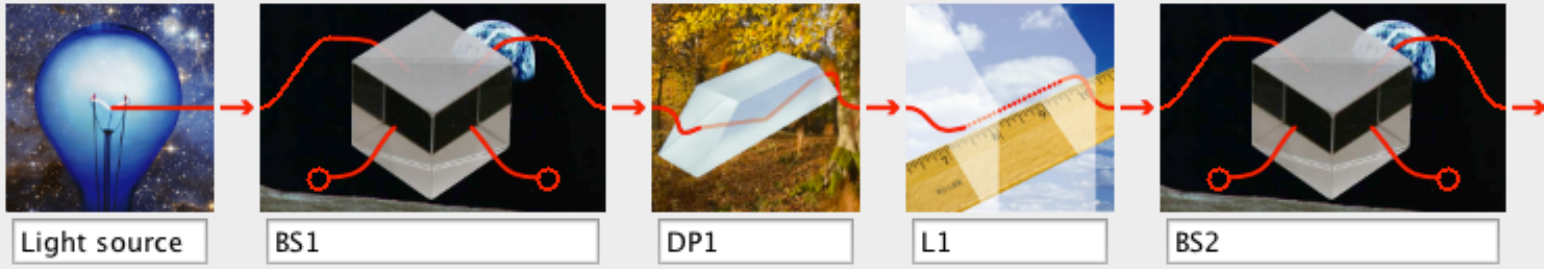} \end{center}
\caption{\label{OAM-sorter-top-arm-figure}\emph{Young TIM}'s representation of the upper arm of the interferometer in Fig.\ \ref{OAM-sorter-figure}.}
\end{figure*}

We start building up the upper part of the optical system represented in Fig.\ \ref{OAM-sorter-figure}(b) by
inserting, one behind the other, beam splitter BS$_1$ and setting its name to ``BS1'';
Dove prism DP$_1$, setting its name to ``DP1'' and its rotation angle to $90^\circ$;
a \texttt{Distance} element that represents a specific distance $L_1$, for example $L_1 = 1\,$m (name ``L1'', $\Delta z = 1$\,m;
we have inserted the distance representing the entire optical path length behind the Dove prism, as the Dove prism simply mirrors the beam, and it does not matter where it does that);
and beam splitter BS$_2$ (name ``BS2'').
The panel at the top of \emph{Young TIM}'s screen area should now display a graphical representation of the part of the optical system we have entered so far (Fig.\ \ref{OAM-sorter-top-arm-figure}).

We still need to add the components representing the interferometer's lower arm (Fig.\ \ref{OAM-sorter-figure}(b)).
\emph{Young TIM}'s graphical representation of the optical system at the top of its screen area always shows only a linear series of components in the optical system, and light always travels through it from left to right.
In Fig.\ \ref{OAM-sorter-top-arm-figure}, for example, the selected series of components corresponds to the upper arm of the interferometer in Fig.\ \ref{OAM-sorter-figure}; specifically, it starts with the beam splitter BS$_1$, shows the components connected to its upper output, namely DP$_1$, the distance element, and BS$_2$, entered through its upper input.
First, we show the series of components connected to the lower output of BS$_1$ by clicking on the red circle shown at the end of the red line corresponding to this output.
Currently, no components are connected to this output, so we connect to it Dove prism DP$_2$ (name ``DP2'', rotation angle $0^\circ$) and plane $P$.
Next, we need to connect elements to the lower input of BS$_2$.
Do do this, we first need to show again a series of components that contains BS$_2$, which can be achieved by clicking on the red circle corresponding to the upper output of BS$_1$.
We can then show the series of components connected to BS$_2$'s lower input by clicking on the red circle corresponding to that input.
We connect to it a distance element representing the distance $L_2$ (name ``L2'', $\Delta z = 1$\,m), and to this an image of the plane $P$ (component type \texttt{Image of plane}, Object ``P'').
The link between the plane $P$ and its image completes the lower arm of the interferometer shown in Fig.~\ref{OAM-sorter-figure}.

The interferometer can now be put to work.
For example, we can attach a Laguerre-Gaussian beam to one of the inputs of BS$_1$ and a plane to each output of BS$_2$, and, after pressing the ``Go!'' button, observe how the interferometer routes a Laguerre-Gaussian beam into one output if the $l$ index of the Laguerre-Gaussian beam (the ``topological charge'' of the optical vortex at the centre of the beam) is even and the other output if $l$ is odd.

It is worth noting the following points:
\begin{enumerate}
\item In most cases, the names given to optical components are there simply to help the user navigate through the optical system.
The name of a plane can be more important: when setting up an image of a plane, the plane that is being imaged is selected from a drop-down list of the names of all planes present in the optical system.

\item If an output from one component is linked to an input of another component, then the light-beam cross-section that leaves the former component is precisely the light-beam cross-section that arrives at the latter component.
Propagation through a distance is represented by a special ``optical component'' of type \texttt{distance}, which simulates diffraction.

\item There is no need to represent planar mirrors, because they change not only the propagation direction of the beam, but also the direction of the optical axis.
This is of no significance to our simulation as \emph{Young TIM} always simulates propagation along the optical axis.

\item For a number of reasons (see appendix \ref{plane-appendix}), \emph{Young TIM} does not allow elements to be linked more than once, at least not directly\footnote{Such a data structure goes by several technical names, including polytree, singly connected network, and directed acyclic graph.}.
This means that the two beam splitters, which are already linked by one arm of the interferometer, cannot be linked again via the other arm.
This means that the second arm has to include an indirect link.
In the representation shown in Fig.\ \ref{OAM-sorter-figure}, this is formed by linking the output of the (bottom) Dove prism (which is linked to output 2 of the left beam splitter via a distance object) to a special optical element that defines a plane ($A$), and linking another special optical element that defines a 1:1 image of that same plane ($A^\prime$), again via a distance object, to input 2 of the right beam splitter.
The advantages of introducing such indirect links are discussed in appendix \ref{plane-appendix}.

\end{enumerate}

\section{Suggested numerical experiments\label{exercises-section}}

\noindent
Here we simply list a few of the numerical experiments students can perform with \emph{Young TIM}, alongside a real experiment or not.
Running a numerical simulation alongside an experiment has a number of advantages; for example, it provides an easy (and cheap) way to try out changes to the experimental setup, and it allows easy access to information that is cumbersome to measure experimentally, for example a beam's phase cross-section in a transverse plane (which would require interferometry in an experiment) or its (phase or intensity) cross sections in a longitudinal plane (which would require scanning).

\subsection{Near- and far-field diffraction}

\noindent
The double-slit experiment that is \emph{Young TIM}'s default optical system can easily be modified, and we outline here a number of ways.

The aperture type can be changed, for example to a single slit or a grating.
Inserting a plane immediately behind the aperture enables near-field (Fresnel) diffraction to be investigated, either in the \texttt{XY plane} tab by changing the additional propagation distance to plot the field in a different transverse plane, or in the \texttt{Z plane} tab by plotting the field in a longitudinal plane.
It is easy to see that making the aperture larger makes the far-field (Fraunhofer) diffraction pattern smaller and vice versa.
In multiple-slit apertures (such as the double slit or grating), it is also easy to disentangle the role of the slit width: by changing the aperture to a single slit of the same width it becomes clear that the single-slit diffraction pattern is the envelope for the multiple-slit diffraction pattern.

\subsection{Formation of laser modes}

\noindent
In our undergraduate labs, the students perform an experiment on laser modes.
In it, they adjust the orientation of the output coupler of the stable canonical cavity of a HeNe laser to select different Hermite-Gaussian modes.
They then also convert these into Laguerre-Gaussian modes (using a cylindrical-lens mode converter \cite{Beijersbergen-et-al-1993}; see Fig.\ \ref{LG-anaglyph-figure}(a)) and investigate their phase structure by interfering the beam with its own mirror image in a Mach-Zehnder interferometer with a Dove prism in one arm~\cite{Padgett-et-al-1996}.

A numerical simulation of this experiment alongside the actual experiment can help the students to improve their understanding of the experiment.
Most notably, they can set up a laser cavity and watch the lowest-loss eigenmode form over successive round trips (the so-called Fox-Li method for finding the lowest-loss eigenmode \cite{Siegman-1986-Fox-Li}).
They can also directly see the phase structure of all beams involved and understand the origin of the radial interference fringes (``spokes'') in the interference experiment, and why they are bent (because the beam that passes through the arm of the interferometer that contains the Dove prism travels a greater optical path length, and so its wavefront curvature is greater).
Measuring and visualising the phase structure of a beam in a real experiment is cumbersome in our undergraduate labs; watching the beam evolve over successive round trips requires heroic effort in \emph{any} lab.

\begin{figure}
\begin{center}
\includegraphics{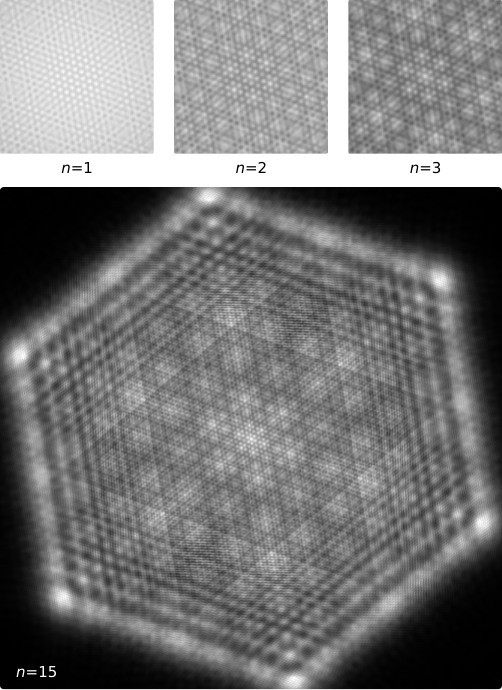}
\end{center}
\caption{\label{laser-mode-formation-figure}Laser-mode formation over successive round trips.
In this example, a fractal laser mode \cite{Courtial-Padgett-2000b,Watterson-et-al-2003} forms in an unstable canonical resonator with a hexagonal intra-cavity aperture and round-trip magnification 3.
The beam's intensity cross-section in the magnified self-conjugate plane after $n=1$, 2, 3 and 15 round trips through the resonator is shown, starting with a Gaussian beam.
For $n=1$, 2 and 3, the picture shows only the centre of the beam.
For $n = 15$, the full beam is shown.}
\end{figure}

Interesting digressions on the topic of laser-mode formation, all of which can be simulated with \emph{Young TIM}, include the formation of fractal modes in unstable resonators (Fig.\ \ref{laser-mode-formation-figure}) \cite{Courtial-Padgett-2000b,Watterson-et-al-2003}, and imaging in stable canonical resonators, which leads to the structural stability of the eigenmodes \cite{Forrester-et-al-2002}.

\subsection{\label{holography-subsection}Holography}

\noindent
One area that can be explored with \emph{Young TIM} is basic holography.
There are many aspects that merit such an exploration; we use it to encourage active experimentation with computer-generated holograms, accompanying a lecture on this subject.

\begin{figure}
\begin{center} \includegraphics{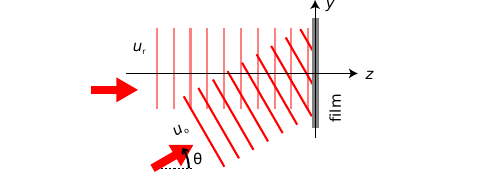} \end{center}
\caption{\label{hologram-recording-figure}Recording of the hologram of a plane wave.}
\end{figure}

Imagine two plane waves of equal amplitude and wavelength, $\lambda$, but different directions, incident on the $z=0$ plane.
We restrict ourselves to waves whose direction lies in the $(y,z)$ plane (it is a useful exercise for students to consider this problem generally).
The first wave, the reference wave, is travelling in the positive $z$ direction;
the second wave, the object wave, is travelling at an angle $\theta$ with respect to the $z$ axis (Fig.\ \ref{hologram-recording-figure}).
The complex amplitudes describing their spatial parts (i.e.\ solutions of the Helmholtz equation) are respectively
\begin{equation}
u_\mathrm{r}(y,z) = \exp(\rmi k z),
\end{equation}
and
\begin{equation}
u_\mathrm{o}(y,z) = \exp[ \rmi (y k \sin \theta + z k \cos \theta + \varphi) ],
\label{object-wave-equation}
\end{equation}
where $k = 2 \pi/\lambda$ and $\varphi$ is an arbitrary phase offset of the object wave with respect to the reference wave.

A film at $z = 0$ would record the following intensity pattern:
\begin{align}
I(y)
&= |u_\mathrm{r}(y,0) + u_\mathrm{o}(y,0)|^2
= 2 + 2 \cos (y k \sin \theta + \varphi).
\label{hologram-intensity-equation}
\end{align}
The film is developed into a hologram: a transparency with a spatially-varying amplitude transmission profile proportional to $I(y)$ as calculated in Eqn (\ref{hologram-intensity-equation}). For convenience, we choose here the proportionality factor to be $1/4$ so that the transmission profile takes values between 0 and 1.

\begin{figure}
\begin{center} \includegraphics{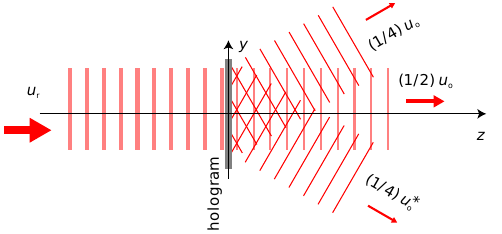} \end{center}
\caption{\label{hologram-reconstruction-figure}Holographic reconstruction of the object wave.}
\end{figure}

Now remove the film and the object wave and place the hologram in the $z=0$ plane.
The reference wave is still there, illuminating the hologram.
Then the field immediately behind the hologram is
\begin{equation}
u(y,0) = \frac{1 + \cos(y k \sin \theta + \varphi)}{2}.
\label{field-behind-hologram-equation}
\end{equation}
This field can be written in the form
\begin{equation}
\begin{split}
u(y,0)
&= \frac{1}{2} u_\mathrm{r}(y,0) + \frac{1}{4} u_\mathrm{o}(y, 0) + \frac{1}{4} u_\mathrm{o}^\ast(y,0),
\end{split}
\label{hologram-reconstruction-equation}
\end{equation}
which is 
the sum of the following three plane waves (see Fig.\ \ref{hologram-reconstruction-figure}):
\begin{enumerate}
\item $(1/2) u_\mathrm{r}(y,z)$, which is a reduced-amplitude reference wave;
\item $(1/4) u_\mathrm{o}(y,z)$, which is a reduced-amplitude object wave;
\item $(1/4) u_\mathrm{o}^\ast(y,z)$, which is a reduced-amplitude mirror image of the object wave (travelling at an angle $-\theta$ with respect to the $z$ axis).
\end{enumerate}
An insightful exercise for students is to work out the effect on the reconstructed object wave of changing the modulation depth, direction, spatial frequency, and sideways shift of the sinusoidal pattern described by Eqn (\ref{field-behind-hologram-equation}).

\begin{figure}
\begin{center} \includegraphics[width=0.5 \columnwidth]{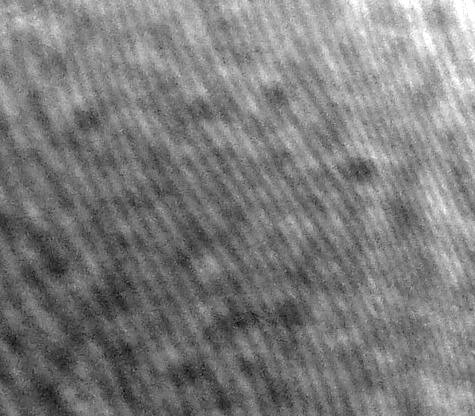} \end{center}
\caption{\label{hologram-pattern-figure}Photo of a small part of a hologram.  (From \url{http://en.wikipedia.org/wiki/Holography}.)} 
\end{figure}

By recognising that locally any wave is a plane wave, our calculation can be generalised to non-planar object and reference waves.
The interference pattern in a plane between a general object wave and a general reference wave (both with wavelength $\lambda$) would locally be a sinusoidal grating like the one described in Eqn (\ref{hologram-intensity-equation}) (Fig.\ \ref{hologram-pattern-figure}); modulation depth, direction and spatial frequency of the grating would change with position.

\begin{figure}
\begin{center} \includegraphics{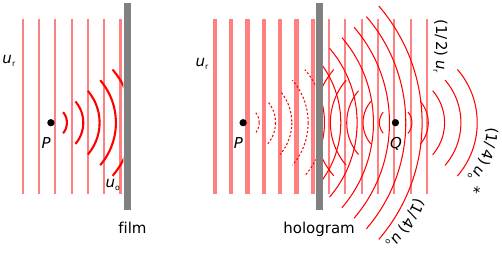} \end{center}
\caption{\label{hologram-of-point-scatterer-figure}Holographic recording (left) and reconstruction (right) of the object wave due to a point scatterer at $P$.
In the reconstruction, fields proportional to the original object wave $u_\mathrm{o}$ and its complex conjugate, $u_\mathrm{o}^\ast$, appear.
The latter gives rise to an image of $P$ on the opposite side of the hologram, namely at position $Q$.}
\end{figure}

\begin{figure}
\begin{center} \includegraphics{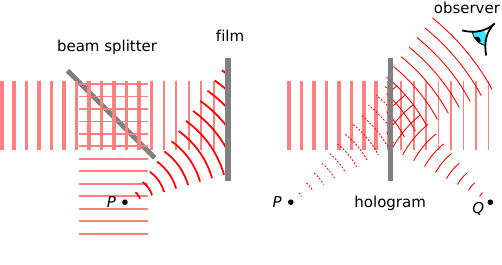} \end{center}
\caption{\label{off-axis-holography-figure}Off-axis holography.
During recording (left), the object is placed to the side of the reference wave that hits the hologram.
Right: After some propagation distance, the reconstructed object wave no longer overlaps with the other waves.  From a suitable viewing position, an observer would only see the object wave, and therefore see only an image of the object and no other images.}
\end{figure}

If an object is reconstructed, the mirror image of the object wave causes a pseudoscopic image of the object (Fig.\ \ref{hologram-of-point-scatterer-figure}).
This additional image can be effectively removed through off-axis holography \cite{Leith-Upatnieks-1962} (Fig.\ \ref{off-axis-holography-figure}).
This works by placing the object to one side, such that the object wave reaches the film at an angle.
When the wave is reconstructed from the hologram, an observer whose eyes intersect only the original wave (and neither its complex conjugate nor the reference wave) sees only the reconstructed object.

\begin{figure}
\begin{center}
\includegraphics{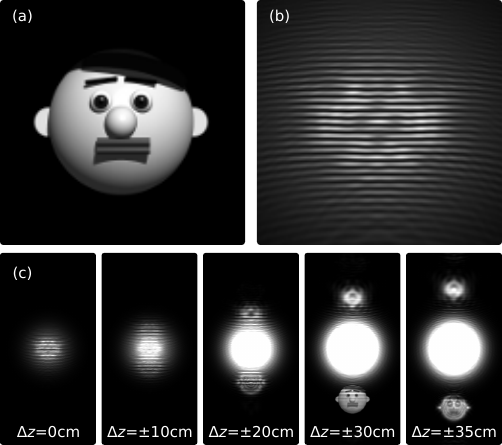}
\end{center}
\caption{\label{holography-figure}Holography with \emph{Young TIM}.
(a)~Object; 
(b)~intensity hologram of the object positioned 30\,cm in front of the film/hologram at an angle $0.5^\circ$ below the optical axis; (c)~reconstruction a distance $\Delta z$ behind the hologram plane.
The reference beam was Gaussian beam travelling along the optical axis and with its waist (waist size 2\,mm) positioned in the film/hologram plane.
In the image for $\Delta z = \pm 30\,\mathrm{cm}$, the reconstructed object wave and its complex conjugate can be seen below and above the central reference beam.
All beams were simulated on a $512 \times 512$ amplitude matrix representing a physical area 1\,cm$\times$1\,cm.
In (a) and (b), only the central 2.5\,mm$\times$2.5\,mm are shown.
In (c), the central 5\,mm$\times$1\,cm are shown and the images are overexposed by different amounts to show up the reconstructed object wave and its complex conjugate.
}
\end{figure}

Fig.\ \ref{holography-figure} demonstrates \emph{Young TIM}'s capabilities of simulating off-axis holography.
The object is a grayscale image (Fig.\ \ref{holography-figure}(a)) positioned in the ``object plane'', a transverse plane 30\,cm in front of the film/hologram (more precisely, in the object plane the intensity cross-section of the object wave is in the shape of an image of a head) and not centred on the axis of the reference wave (so that we simulate off-axis holography).
The reference wave is a Gaussian beam whose beam axis coincides with the optical axis; its waist (of waist size 2\,mm) is positioned in the film/hologram plane.
The object and reference waves must be represented such that they can be meaningfully combined, i.e.\ they must use the same amplitude-matrix size (in our case 512$\times$512), the same represented physical area (in our case 1\,cm$\times$1\,cm), and the same wavelength (in our case 632.8\,nm).
We ensure that the maximum intensity in the reference wave is the same as that in the object wave by passing both through \texttt{Neutral-density filter} elements configured to make the maximum intensity in both beams 1.
We combine the object wave and reference wave by passing them into the two input arms of a \texttt{Beam splitter} optical component, and we observe the resulting interference pattern as the intensity cross-section in a \texttt{Plane} optical element placed immediately behind one of the beam-splitter outputs (Fig.\ \ref{holography-figure}(b)).
We save this intensity cross-section to a bitmap file, which becomes the source of the hologram pattern for the reconstruction

We create the object wave with the desired properties, and simultaneously with good overlap with the reference wave in the film/hologram plane, as follows.
We initiate the beam with a \texttt{Light source} element that produces a beam of beam type \texttt{Beam profile from bitmap}.
This element represents the object wave in the object plane, travelling coaxially with the reference wave (we later add an element that makes it travel at an angle to the reference wave).
The bitmap image that determines the intensity and phase of the beam is an image file we prepared earlier\footnote{We used raytracer TIM \cite{Lambert-et-al-2012,Oxburgh-et-al-2014} to create an image of a head, saved it as BMP file, and then used freeware GIMP to convert it into a grayscale image and crop/resample it to size 512$\times$512 pixels.
Note that the head itself only covers the centre of the image.}, and this file can be selected in the \texttt{Light source}'s control panel.
We then propagate this beam from the object plane to the hologram plane, i.e.\ through a distance 30\,cm (using a \texttt{Distance} element), and then angle it with respect to the optical axis by passing it through a \texttt{Wedge} element (deflection angle in the $(y,z)$ plane is $0.5^\circ$).

To calculate the beam behind the hologram, we start with the reference beam (i.e.\ Gaussian beam of waist size 2\,mm and wavelength 632.8\,nm, represented on a 512$\times$512 amplitude matrix representing an area 1\,cm$\times$1\,cm) and pass it through an optical component of type \texttt{Hologram (from bitmap)}, which allows the user to select the previously saved bitmap file as the hologram pattern.
We then pass the beam directly into a \texttt{Plane} element.
In the \texttt{Plane}'s \texttt{XY plane} tab, the beam can then be propagated into transverse planes in front of and behind the hologram plane (Fig.\ \ref{holography-figure}(c)), specifically the object plane (30\,cm in front of the hologram) and its mirror image with respect to the hologram plane (i.e.\ a transverse plane 30\,cm behind the hologram).

\section{\label{anaglyphs-section}A slightly special power: creating anaglyphs that simulate the view when looking into the beam}

\noindent
Looking directly into the beam from a laser is not usually recommended because of safety concerns.
Of course, laser beams can be made safe, for example by limiting the power in the beam to be in class 1, but nevertheless:  laser physicists generally have no idea what it looks like to look directly into different laser beams, just like chemists generally no longer have any idea what their chemicals, particularly the toxic variety, taste like\footnote{The pharmaceutical chemist Carl Wilhelm Scheele (1742-1786), for example, who is now considered to be the first to discover oxygen and a number of other chemical elements and compounds, had one notable shortcoming, namely ``a curious insistence on tasting a little of everything he worked with, including such notoriously disagreeable substances as mercury, prussic acid (another of his discoveries) and hydrocyanic acid --- a compound so famously poisonous that 150 years later Erwin Schr{\"o}dinger chose it as his toxin of choice in a famous thought experiment [...].  In 1786, aged just forty-three, he was found dead at his workbench surrounded by an array of toxic chemicals, any one of which could have accounted for the stunned and terminal look on his face.''~\cite{Bryson-2003}}.

\begin{figure}
\begin{center} \includegraphics[width=\columnwidth]{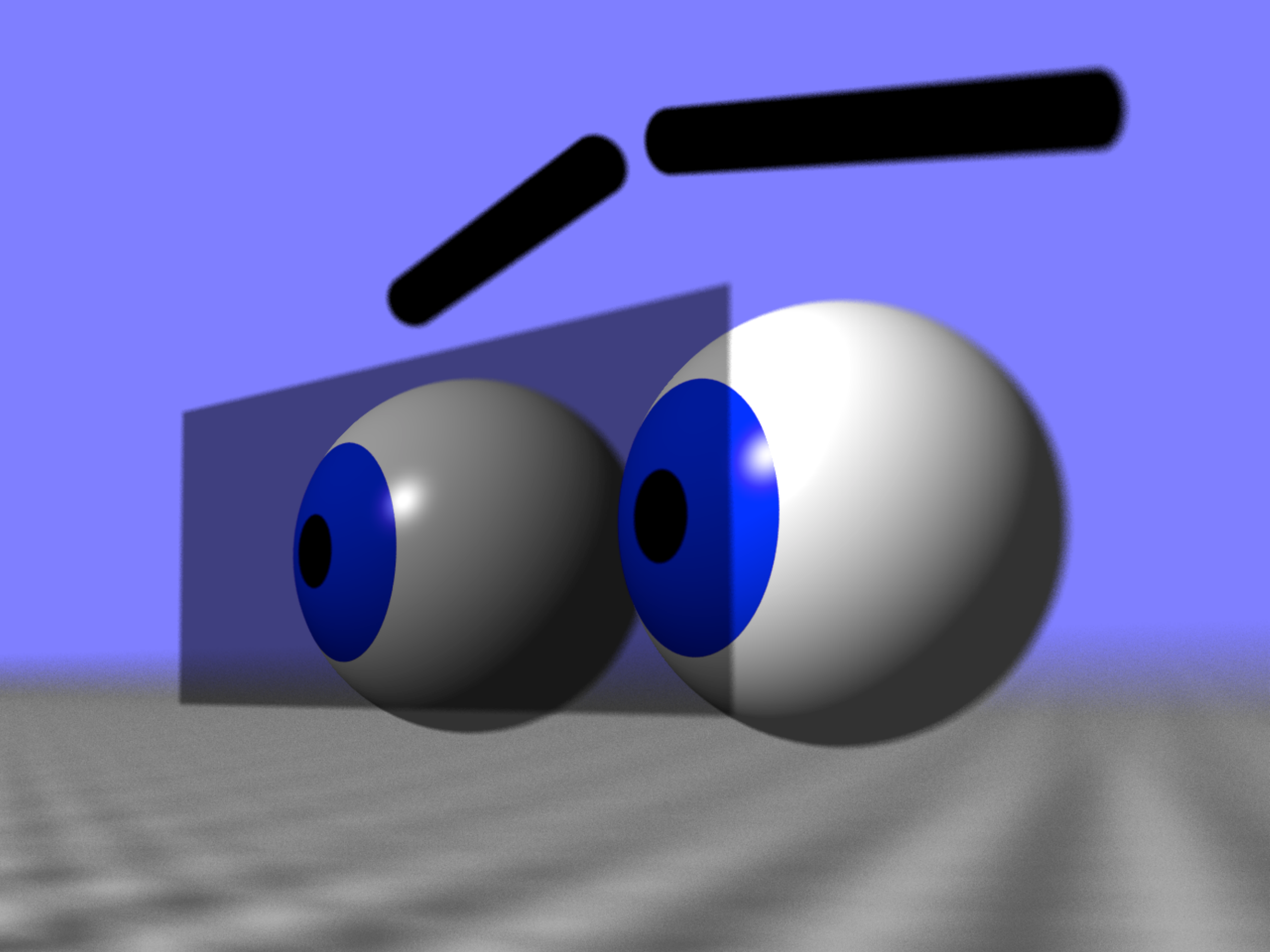} \end{center}
\caption{\label{eyes-and-wave-front-figure}Wave front (shaded rectangle directly in front of eyes) reaching a pair of eyes.
}
\end{figure}

We have built into \emph{Young TIM} the ability to create anaglyphs of the three-dimensional view a binocular observer would be experiencing when looking directly into the beam.
This is one of \emph{Young TIM}'s main ``slightly special'' powers.
Fig.\ \ref{eyes-and-wave-front-figure} is an artist's impression of a wave front reaching a pair of eyes.
Clearly, the left and right eyes intercept different parts of the beam, namely the part passing through either the left eye's pupil or the right eye's pupil\footnote{By default, \emph{Young TIM} uses for the interpupillary distance (IPD, or simply Pupillary Distance, PD) 63\,mm, which is the approximate mean value of a standard dataset \cite{Dodgson-2004} (which shows amongst other things that, like fridges, IPD is bigger in America than elsewhere).
Pupil diameters range considerably with lighting conditions and age (pupil size decreases with age for similar lighting conditions); in Ref.\ \cite{Winn-et-al-1994}, for example, a range of 2\,mm to 9\,mm was measured.
By default, \emph{Young TIM} uses a pupil radius of 2\,mm (i.e.\ a diameter of 4\,mm).}.
Each different beam part then passes through the corresponding eye's lens and propagates to the retina.
To create an anaglyph, all that is required is to display the intensity on the left eye's retina in red, that on the right eye's retina in blue, combine them in a single image and view this image with standard red/blue anaglyph goggles.

The way \emph{Young TIM} finds the shape of the beam on the retina of each eye after aperturing due to the pupil is as follows.
Fourier transforming the apertured beam is equivalent to calculating the beam in the back focal plane of the eye's lens, which in turn is an image of the beam in a plane an infinite distance in front of the lens.
Ray-optically, Fourier-transforming means focussing each set of parallel rays into a point.
A lens of focal length $f$ makes rays that originated from its front focal plane parallel, and so simulating passage through such a lens before Fourier transforming gives an image of the beam in a plane a distance $f$ in front of the lens.
Two lenses, one immediately behind the other, are equivalent to a single lens whose focussing power (i.e.\ inverse focal length) is the sum of the focussing powers of the individual lenses; the addition of a lens is equivalent to a change in focussing power of the (one) lens, which is precisely how the human eye focusses to different distances.


\begin{figure}
\begin{center} \includegraphics{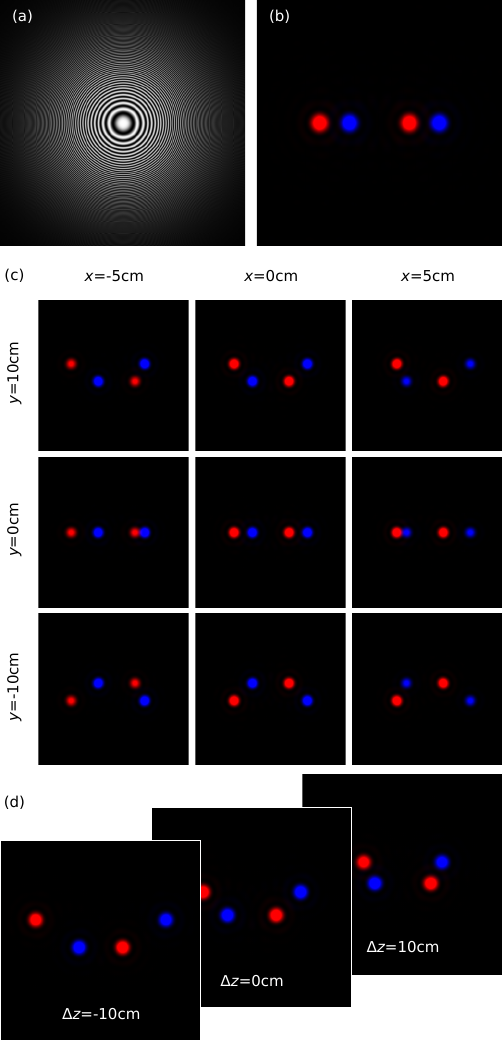} \end{center}
\caption{\label{anaglyphs-figure}(a)~Intensity in the $(x,y)$ plane (phase not shown) of a superposition of two spherical waves and corresponding anaglyphs (b--d).
(b)~Standard parameters (pupil radius 2\,mm, interpupillary distance 63\,mm, pupils centred at $(x, y) = (0, 0)$, additional propagation distance $\Delta z = 0$, focussing distance 1\,km;
(c)~different $x$ and $y$ coordinates of the midpoint between the pupils (otherwise like (b));
(d)~different additional propagation distances $\Delta z$ (otherwise like (b) with $x=0$ and $y=10\,\mathrm{cm}$.
The figure was calculated using an amplitude matrix of $512 \times 512$ elements, representing a physical area of size 10\,cm$\times$10\,cm.
The spherical waves were centred at positions $(0, 0, 80\,\mathrm{m})$ and $(0, 0, -40\,\mathrm{m})$, respectively.}
\end{figure}

Fig.\ \ref{anaglyphs-figure} shows anaglyphs for a simple beam, namely a superposition of two spherical waves.
Frame (b) shows the anaglyph calculated for standard parameters.
The anaglyph represents two diffraction-limited spots, one behind the paper plane, the other in front, but this is by no means obvious, as the anaglyph is ambiguous because all spots lie on the same horizontal line and it is not clear which ones correspond to each other.
Frames (c) and (d) demonstrate the use of the additional parameters to remove this ambiguity.
In frame (c), the pupils are moved transversally in the beam, in frame (d), the additional propagation distance $\Delta z$ is varied.
The effect that of parallax associated with different viewing positions, moved transversally in (c) and longitudinally in (d).
Note that these parameters can be altered interactively in \emph{Young TIM}, and the image changes almost instantaneously (depending on resolution and computer speed), and so the user can change the viewing position interactively and in real time.

\begin{figure}
\begin{center}
\includegraphics{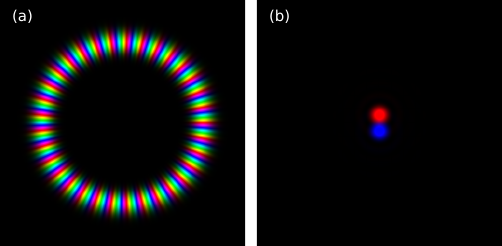}
\end{center}
\caption{\label{LG-anaglyph-figure}Phase \& intensity cross-section~(a) and corresponding anaglyph~(b) of a Laguerre-Gaussian (LG) mode.
The figure was calculated using an amplitude matrix of $512 \times 512$ elements, representing a physical area of size 10\,cm$\times$10\,cm.
The LG mode has a waist size $w_0 = 8\,\mathrm{mm}$ and mode indices $l = 33$, $p = 0$.
The anaglyph was calculated with standard anaglyph parameters.}
\end{figure}

Incidentally, anaglyphs cannot always be interpreted easily as originating from objects such as the point light sources at the centre of the spherical waves in the previous example.
One example are anaglyphs of optical vortices (also known as phase singularities), which are common features of random waves \cite{Berry-Dennis-2000a}.
Fig.\ \ref{LG-anaglyph-figure} shows the anaglyph for a Laguerre-Gaussian (LG) beam, which contains a central, high-order, optical vortex.
It can be seen that each eye sees one spot of light, but the spots are not shifted horizontally, which might be expected due to the parallax caused by the horizontal separation of the eyes, but vertically.
Such a displacement is not consistent with a (real) object; in fact, light-ray fields such as those in LG beams can be described mathematically as originating from a point light source at a \emph{complex} distance~\cite{Courtial-et-al-2012}.

%
%

\section{\label{artefacts-section}The importance of understanding the effect of optical components}

\noindent
Experimenting with complex simulation software makes our students aware of the shortcomings and limitations of simulations.
It is important to spot such artefacts when they occur, so that they can then dealt with and ideally avoided.
For both detection and avoidance of artefacts it is often important to have a reasonable understanding of how the simulation works.
We use here as examples artefacts introduced by simulated free-space propagation, caused by the algorithm's assumption of periodicity in both position and momentum space.
These artefacts show up if the parameters are not chosen appropriately~\cite{Courtial-1999a-beam-propagation-problems}.


By default, \emph{Young TIM} represents the beam amplitude on a grid of points that covers a rectangle in a transverse plane.
On propagation from a rectangle in one transverse plane to the equivalent rectangle in another transverse plane, the beam changes.
What happens if, in the new transverse plane, it does no longer lie in the represented rectangle?
A reasonable assumption would be that it simply disappears.

\begin{figure}
\begin{center} \includegraphics{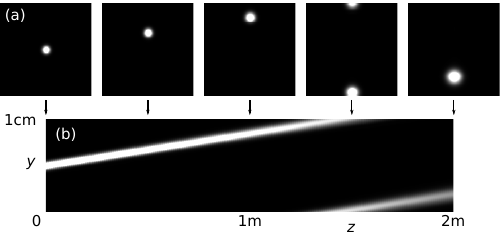} \end{center}
\caption{\label{space-periodicity-figure}Periodicity in position space.
The images show intensity-cross-sections through a Gaussian beam travelling at an angle $0.2^\circ$ in the $yz$ plane with respect to the $z$ axis.
An inclined Gaussian beam can be created by passing the beam from a light source of beam type \texttt{Gaussian beam} (which travels parallel to the optical axis) through a wedge with a deflection angle (yz projection) of $0.2^\circ$.
The Gaussian beam had a waist size of 0.5\,mm and a wavelength $\lambda = 632.8$\,nm.
The simulation was performed on a $128 \times 128$ grid representing a physical square of size 1\,cm$\times$1\,cm.
(a)~$(x,y)$ cross sections at propagation distances $z = 0$, 0.5\,m, 1\,m, 1.5\,m, and 2\,m.
The full 1\,cm$\times$1\,cm simulation area is shown.
(b)~$(y,z)$ cross section shown from $z = 0$ to $z=2$\,m.
The area shown is 1\,cm in the $y$ direction and 2\,m in the $z$ direction, so the plot is compressed in the horizontal direction, which is why the angle of the beam with the horizontal is significantly greater than $0.2^\circ$.}
\end{figure}

Fig.\ \ref{space-periodicity-figure} shows what actually happens.
It shows the simulated intensity cross-section of a Gaussian beam in various transverse planes.
The axis of the Gaussian beam is inclined with respect to the $z$ axis such that, on propagation through positive $z$ distances, the beam's intensity cross-section moves up (it also gets wider).
The figure shows that, as the beam leaves one side of the represented rectangle, it re-appears on the opposite side.
Mathematically, this is because the discrete Fourier transform (DFT), which \emph{Young TIM} uses when calculating the beam propagation, implicitly assumes periodic boundary conditions:
what \emph{Young TIM} is actually simulating is one unit cell in an infinite lattice, periodic in the $x$ and $y$ directions, and when light leaves the unit cell to one side, light from the neighbouring unit cell on the opposite side also leaves that neighbouring unit cell and enters the cell simulated by \emph{Young TIM}.




\begin{figure}
\begin{center} \includegraphics{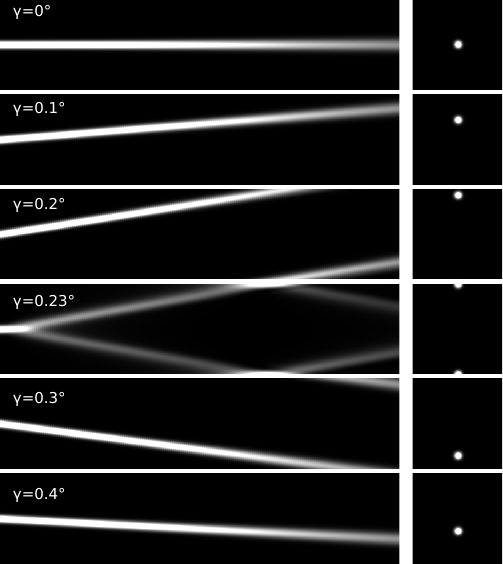} \end{center}
\caption{\label{momentum-space-periodicity-figure}Periodicity in position and momentum space.
The plots in the left-hand column are of the same type as Fig.\ \ref{space-periodicity-figure}(b) (in fact, the plot for $\gamma = 0.2^\circ$ is the same plot), for the same beam, but with phase gradients that correspond to different angles $\gamma$ in the $yz$ plane with respect to the $z$ axis.
From top to bottom, the angle $\gamma$ is increased from $0^\circ$ to $0.4^\circ$.
Note that, like in Fig.\ \ref{space-periodicity-figure}, the plots are compressed in the horizontal direction.
The right-hand column shows the corresponding distributions in the represented area of transverse momentum space.
The horizontal and vertical axis show wavenumber in the horizontal and vertical direction, $k_x$ and $k_y$, respectively.}
\end{figure}

In addition to periodicity in the transverse directions, $x$ and $y$, the discrete Fourier transform also assumes periodicity in transverse momentum space (or wavenumber space, Fourier space).
Fig.\ \ref{momentum-space-periodicity-figure} illustrates this.
It shows simulations of beams similar to that shown in Fig.\ \ref{space-periodicity-figure}, but travelling at different angles $\gamma$ with the $z$ axis.
The intensity cross-sections in the $(y, z)$ plane show that, for small angles $\gamma$, the beam propagates as expected, namely in a straight line that is initially horizontal ($\gamma = 0^\circ$) and then slightly inclined upwards ($\gamma = 0.1^\circ$; note that the cross-section plots are compressed in the horizontal direction).
The plot for $\gamma = 0.2^\circ$ is the same as that shown in Fig.\ \ref{space-periodicity-figure}(b), and illustrates again the periodicity of position space.
As $\gamma$ is increased to around $\gamma = 0.23^\circ$, the beam first splits into two beams inclined with respect to the horizontal axis by angles of equal magnitude but opposite sign.
Upon further increase of $\gamma$, the beam inclined upwards disappears and only the downwards-inclined beam remains.

The explanation for the appearance of a beam with an inclination of equal magnitude but opposite sign lies in the limited size and periodicity of the represented area of momentum space (also known as $k$ space), which is also shown in Fig.\ \ref{momentum-space-periodicity-figure}.
More precisely, \emph{Young TIM} calculates the propagation of waves using a Fourier algorithm (see App.\ \ref{distance-subsection}) 
that represents the beam cross-section by its discrete Fourier Transform (DFT), each element of which corresponds to a uniform plane wave with specific horizontal and vertical wavenumbers, $k_x$ and $k_y$, and the plane wave represented by the DFT corresponds to a rectangle in the $(k_x, k_y)$ plane, centered on $(k_x, k_y) = (0, 0)$.
At $\gamma = 0^\circ$, the momentum distribution is a centred Gaussian, and as $\gamma$ is increased it moves upwards.
At approximately $\gamma = 0.23^\circ$, the Gaussian is so high that half of it has disappeared out of the represented area of momentum space, but because of the periodicity has re-appeared at the bottom.
At this point, the phase gradient has reached the Nyquist limit, which corresponds to a phase difference of $\pi$ between neighbouring points on the grid.
As $\gamma$ is increased further, the Gaussian keeps moving upwards.






A simple test for the problems described above is to run the simulation again with a larger represented area, or a larger amplitude-matrix size, or both.
Any change in the outcome of the simulation is indicative of a problem.


%

\section*{Conclusions}

\noindent
We initially started to write \emph{Young TIM} as a research tool, but quickly realised its usefulness in teaching of wave optics.
Now \emph{Young TIM} is firmly embedded in a number of experiments in our undergraduate laboratory.
In addition to the benefits for performing additional experiments when time, budget, or even the laws of physics would otherwise not allow this, we believe this experience of complex simulation software run alongside the experiment gives students a more realistic experience of physics research.
Additionally, it might find use in encouraging students to prepare before starting an experiment, possibly in combination with a pre-experiment formative assessment (such as Just-in-Time Teaching (JiTT) \cite{Novak-et-al-1999}).
Finally, our students work in pairs, and one should not underestimate the value of running a simulation alongside the experiment for keeping all students busy.




\section*{Acknowledgements}

\noindent
This research received no specific grant from any funding agency in the public, commercial or not-for-profit sectors.

\appendix

\section{\label{components-section}Optical components}

\noindent
This appendix discusses all the optical components available in the interactive version in the order in which they appear in the drop-down menu for inserting new optical components into the optical train.


\subsection{Light source}

\noindent
In order for anything to happen, any series of optical elements needs light to enter it.
In \emph{Young TIM}, this light is provided by special light-source-type components that have no inputs and one output.
They can also have a number of parameters that specify the shape of the output.

The most straightforward of these is the \texttt{Light source} component.
It creates an initial amplitude matrix from its input parameters.
Those parameters describe the beam type (Gaussian, ...) and its measurements (wavelength, waist size, ...), but also the simulation parameters that determine the size of the amplitude matrix and the physical area it represents.

\begin{figure}
\begin{center}
\includegraphics{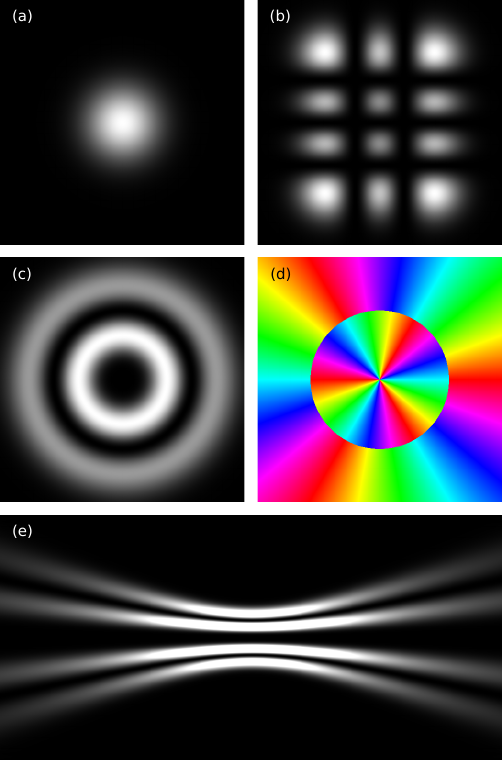}
\end{center}
\caption{\label{beams-figure}Transverse and longitudinal cross sections of different beam types.
(a)~Transverse intensity cross-section of a Gaussian beam.
The waist size is $w_0=2\,\mathrm{mm}$.
(b)~Transverse intensity cross-section of a Hermite-Gaussian beam.
The waist size is $w_0=2\,\mathrm{mm}$, the mode indices are $m = 2$, $n = 3$.
Transverse intensity~(c) and phase (d) cross-section of a Laguerre-Gaussian beam with waist size $w_0 = 2\,\mathrm{mm}$ and mode indices $l=3$, $p = 1$.
In the phase plot, in which the phase is represented as hue, an optical vortex --- a point of uncertain phase, also known as a phase singularity --- is visible in the centre.
(e)~Longitudinal intensity cross section (in the $(x, z)$ plane) of the above Laguerre-Gaussian beam, but with $w_0 = 0.5\,\mathrm{mm}$, plotted from $z=-4\,\mathrm{m}$ to $z=+4\,\mathrm{m}$.
The beam waist can be seen at the centre.
All beams are centred on the origin.
They are represented by an amplitude matrix of size $512\times512$, representing a physical area 1\,cm$\times$1\,cm.}
\end{figure}

A \texttt{Light source} component can create the following beam types:
\begin{description}
\item [Gaussian beam] A Gaussian beam (Fig.\ \ref{beams-figure}(a)) whose beam axis is parallel to the $z$ axis.
Parameters are the waist size, $w_0$, and the $x$ and $y$ coordinates of the beam axis.
The amplitude cross section in the plane of the beam waist is returned.
\item [Hermite-Gaussian beam] A Hermite-Gaussian beam (Fig.\ \ref{beams-figure}(b)), a generalisation of a Gaussian beam with the parameters of a Gaussian beam and additionally the mode indices $m$ and $n$, which specify the number of notes in the horizontal and vertical direction, respectively.
A Hermite-Gaussian beam with mode indices $m=n=0$ is a Gaussian beam.
The amplitude cross section in the plane of the beam waist is returned.
\item [Laguerre-Gaussian beam] A Laguerre-Gaussian beam (Fig.\ \ref{beams-figure}(c,d)) \cite{Allen-et-al-1992}, which is another generalisation of a Gaussian beam with the parameters of a Gaussian beam and additional mode indices $l$ and $p$, which specify the number of full $2\pi$ phase cycles the beam undergoes on any circle around the beam axis and the number of radial nodes (in addition to the central one, which appears when $l \neq 0$).
(Laguerre-Gaussian beams are interesting: for $l \neq 0$, a Laguerre-Gaussian beam forms an optical vortex line along its axis and possesses orbital angular momentum.)
A Laguerre-Gaussian beam with mode indices $l=p=0$ is a Gaussian beam.
The amplitude cross section in the plane of the beam waist is returned.
\item [Uniform plane wave] A uniform plane wave that travels in an arbitrary direction that is given by the transverse wave numbers, $k_x$ and $k_y$.
The amplitude-matrix element corresponding to position $(x, y)$ is set to $\exp(\rmi(k_x x + k_y y))$.
A uniform plane wave with $k_x = k_y = 0$ travels in the direction of the optical axis, i.e.\ the $z$ axis.
\item [Beam profile from bitmap] An arbitrary amplitude cross-section that is created from a bitmap image file.
Each pixel in the bitmap is converted into a complex amplitude of the corresponding element in the amplitude matrix: the pixel's hue becomes the amplitude-matrix-element's phase, its brightness becomes the intensity.
The size of the bitmap in pixels has to be the same as the size of the amplitude matrix in elements (e.g.\ 256$\times$256).
\end{description}

Gaussian beams, Hermite-Gaussian beams, and Laguerre-Gaussian beams are all structurally stable beams, which means that their transverse intensity cross-sections do not change shape on propagation, but they do change size (at the beam waist they are smallest, and brightest --- see Fig.\ \ref{beams-figure}(e)).

\subsection{Image of plane}

\noindent
The output of this ``component'' is simply a copy of the amplitude matrix currently stored in an associated \texttt{Plane} component, the ``object''.
Such a situation is realised in a 1:1 imaging system, when the \texttt{Plane} component represents the object plane and the \texttt{Image of plane} component represents the image plane.
To represent imaging systems with a different magnification, the \texttt{Image of plane} can be followed by a \texttt{Beam telescope}.

In \emph{Young TIM}, this component additionally allows optical systems with some form of optical feedback, for example interferometers or resonators, to be simulated (section \ref{other-optical-systems-section}).

If the \texttt{Plane} component that  has not been initialised, the \texttt{Image of plane} component optionally returns the output from an associated \texttt{Light source} component.

\subsection{Aperture}

\begin{figure}
\begin{center}
\includegraphics{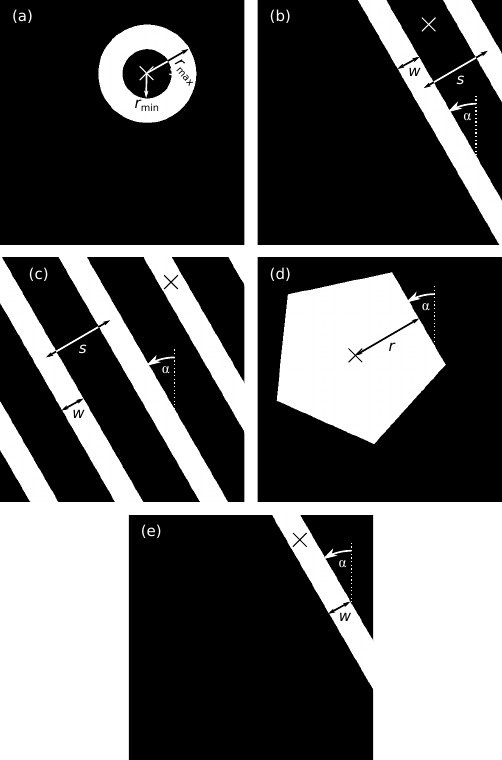}
\end{center}
\caption{\label{apertures-figure}Intensity cross-section immediately behind different apertures.
The intensity cross-section of the incident beam is uniform.
The geometrical meaning of the parameters is indicated in each case; the $\times$ marks the position $(x_\mathrm{centre}, y_\mathrm{centre})$.
(a)~Annular aperture, outer radius $r_\mathrm{outer} = 2\,\mathrm{mm}$, inner radius $r_\mathrm{inner} = 1\,\mathrm{mm}$, centre osition $(x_\mathrm{centre}, y_\mathrm{centre}) = (1\,\mathrm{mm}, 2\,\mathrm{mm})$.
(b)~Double slit, $s = 3\,\mathrm{mm}$, $w = 1\,\mathrm{mm}$, $\alpha = 30^\circ$, $(x_\mathrm{centre}, y_\mathrm{centre}) = (2\,\mathrm{mm}, 4\,\mathrm{mm})$.
(c)~Grating, $s = 3\,\mathrm{mm}$, $w = 1\,\mathrm{mm}$, $\alpha = 30^\circ$, $(x_\mathrm{centre}, y_\mathrm{centre}) = (2\,\mathrm{mm}, 4\,\mathrm{mm})$.
(d)~Polygonal aperture, $N = 5$, $r = 3\,\mathrm{mm}$, $\alpha = 30^\circ$, $(x_\mathrm{centre}, y_\mathrm{centre}) = (-1\,\mathrm{mm}, 1\,\mathrm{mm})$.
(e)~Slit, $w = 1\,\mathrm{mm}$, $\alpha = 30^\circ$, $(x_\mathrm{centre}, y_\mathrm{centre}) = (2\,\mathrm{mm}, 4\,\mathrm{mm})$.
All beams were simulated on a $512\times512$ amplitude matrix representing a physical area 1\,cm$\times$1\,cm.}
\end{figure}

\noindent
The output from an \texttt{Aperture} component is its input with a subset of the elements in the amplitude matrix, namely those elements corresponding to positions where the aperture is absorbing, set to zero.
Which elements these are depends on the aperture type and its parameters.
The following aperture types are currently implemented in \emph{Young TIM}:
\begin{description}
\item [Annular aperture]  An annular aperture (Fig.\ \ref{apertures-figure}(a)) is described by its outer radius, inner radius, and centre position.
\item [Double slit]  A double slit (Fig.\ \ref{apertures-figure}(b)) is described by the centre-to-centre slit separation, the slit width, the angle of the slits with the $y$ direction, and the $x$ and $y$ coordinates of centre of rotated slits (the position described by the selected coordinates will lie half-way between the two slits).
\item [Grating]  A grating (Fig.\ \ref{apertures-figure}(c)) is described by the centre-to-centre slit separation, the slit width, the angle of the slits with the $y$ direction, and the $x$ and $y$ coordinates of centre of rotated slits (the position described by the selected coordinates will lie half-way between two neighbouring slits).
Optionally, the edges can be made ``soft'', i.e.\ the amplitude transmission coefficient $T_\mathrm{a}$ smoothly changes between 0 and 1 like the function
\begin{align}
T_\mathrm{a} = \frac{1}{2} + \frac{1}{2} \sin \left( \pi \frac{u - u_\mathrm{c}}{w} \right),
\end{align}
where $u$ is a coordinate across the edge, which is centred at $u = u_\mathrm{c}$ and of width $w$ (Fig.~\ref{f:softBoundaryTransmissionCoefficient}).
\begin{figure}
\begin{center} \includegraphics{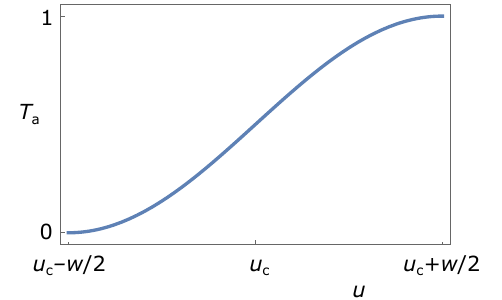} \end{center}
\caption{\label{f:softBoundaryTransmissionCoefficient}Variation of the amplitude transmission coefficient, $T_a$, across a soft edge.
The horizontal axis shows $u$, a coordinate perpendicularly across the edge, which is centred at $u = u_\mathrm{c}$ and has width $w$.}
\end{figure}
\item [Polygonal aperture]  A polygonal aperture (Fig.\ \ref{apertures-figure}(d)) is determined by the number of sides, its radius (i.e.\ closest distance of polygonal sides from centre; the corners are more distant from the centre), the angle of one of the sides with the $y$ direction, and the centre position.
Polygonal apertures are important for fractal lasers 
\cite{Karman-Woerdman-1998,Karman-et-al-1999,Watterson-et-al-2003}.
\item [Slit] The parameters describing a slit (Fig.\ \ref{apertures-figure}(e)) are the slit width, the angle of the slit with the $y$ direction, and the $x$ and $y$ coordinates of the centre of the (rotated) slit.
\item [Absorbing boundary] An absorbing boundary whose width is given in amplitude-matrix elements.
The boundary is a frame surrounding the amplitude matrix.
The frame's edge is soft and of the form shown in Fig.\ \ref{f:softBoundaryTransmissionCoefficient}.
\end{description}

\subsection{Aperture stack}

An \texttt{Aperture stack} represents a stack of apertures, starting immediately in front of the first aperture and finishing immediately behind the last.
The number of apertures, and the separation between neighbouring apertures, can be set.

The thickness of the aperture stack is given by
\begin{align}
\Delta z (n - 1),
\end{align}
where $\Delta z$ is the separation between neighbouring apertures and $n$ is the number of apertures in the stack.
It is also possible to set the separation between neighbouring apertures such that the thickness of the aperture stack reaches a given value.

The type of aperture can be selected from those types available in the \texttt{Aperture} component.

\subsection{Beam expander}

\noindent
The \texttt{Beam expander} component has one parameter: the magnification factor, $M$.
If $M \geq 0$, the output is the same as the input with one difference:
if the input's represented physical area is $w \times h$, the output's represented physical area is $M w \times M h$.
If $M < 0$, the output differs from the input in two ways:
the output's amplitude matrix is the same as the input's amplitude matrix, but rotated by $180^\circ$ around the centre, and the output's $x$ and $y$ components of the represented physical area are those of the input, scaled by a factor $|M|$.

The \texttt{Beam expander} component can represent an idealised beam expander, a laser-physics component.
It can also represent a perfect imaging device with arbitrary magnification.

\subsection{Beam rotator}

\noindent
This component simply rotates the amplitude matrix by an arbitrary angle around the position $(0, 0)$.

Beam rotation can be realised physically with a number of devices \cite{Swift-1972}, for example a pair of rotated Dove prisms (each one mirroring with respect to a different mirror plane, with the combined effect of rotating by twice the angle between the mirror planes around the lines where the mirror planes intersect).
Note that passage through such devices represents a significant optical path length, and so the input and output of the \texttt{Beam rotator} component do not correspond to the beam immediately in front of and immediately behind the rotation device.

\begin{figure}
\begin{center}
\includegraphics{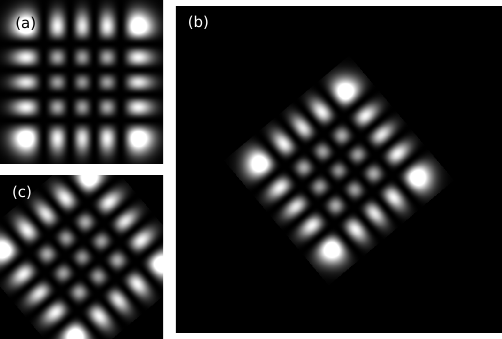}
\end{center}
\caption{\label{beam-rotation-figure}Beam rotation.
(a)~Original beam; (b)~rotated beam on larger amplitude matrix; (c)~rotated beam on amplitude matrix cropped to original size.
The original beam is a Hermite-Gaussian beam with a waist size of 1\,mm and $n = m = 4$, represented on a 256$\times$256 amplitude matrix representing a physical area of 5\,mm$\times$5\,mm.
In (b) and (c), the beam is rotated by $40^\circ$.}
\end{figure}

In fact, \emph{Young TIM} rotates the amplitude matrix not only in the \texttt{Beam rotator} component, but also in a number of other components, specifically in the \texttt{Dove prism} and the \texttt{Dove-prism array}, in order to simulate the effect of rotated components.
In this case, the beam first gets rotated by the negative angle by which the component is rotated, then the component acts on it, and finally it is rotated back into its original orientation.

\begin{figure}
\begin{center}
\includegraphics{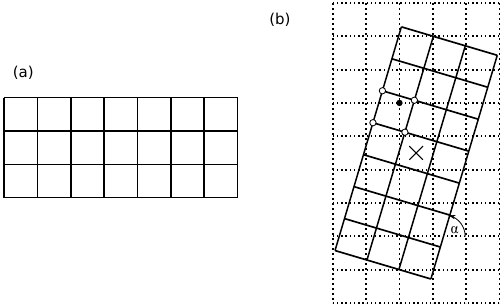}
\end{center}
\caption{\label{rotated-grid-figure}Rotation of a beam cross-section.
(a)~Grid on which the unrotated amplitude cross-section is represented.
The amplitude matrix contains the values of the complex electric field at the intersection points of the grid lines.
(b)~Initial grid after rotation through an angle $\alpha$ around the centre (marked by $\times$).
After beam rotation, the beam is represented on a new grid (dotted lines).
The complex electric field at each intersection point of the lines of the new grid, for example the solid dot indicated in the diagram, is calculated from the complex electric fields at the four nearest points where the lines of the rotated initial grid intersect (white dots).
The new grid generally is bigger than the initial grid to accommodate the rotated initial grid.
In \emph{Young TIM}, the new grid is even bigger than drawn here so that the grid side lengths are powers of two.}
\end{figure}

Every amplitude matrix always represents the complex field on a grid of points that are periodic in the $x$ and $y$ directions.
The amplitude cross-section can be rotated by rotating this grid of points, but it is then generally no longer  periodic in the $x$ and $y$ directions (Fig.\ \ref{rotated-grid-figure}).
\emph{Young TIM} therefore needs to calculate the complex field at a new grid of points that is periodic in the $x$ and $y$ direction.
The complex field at each of these new grid points is calculated through bilinear interpolation \cite{Wikipedia-Bilinear-interpolation} from the complex-field values at the four nearest points on the rotated initial grid (Fig.\ \ref{rotated-grid-figure}(b)).

In general, the new grid has to be bigger than the initial, unrotated, grid in order to accommodate the rotated initial grid (Fig.\ \ref{rotated-grid-figure}(b)).
If the initial grid is of size $N \times M$, we set the size of the new grid to $2 \, \mathrm{max}(N, M) \times 2 \, \mathrm{max}(N, M)$.
This grid can accommodate the rotated initial grid for any rotation angle, and the number of elements in each direction is automatically a power of 2 (as $N$ and $M$ are powers of two), which is a requirement of the DFT algorithm used to propagate the beam (see section \ref{distance-subsection}).

The \texttt{Beam rotator} component gives the user the choice to return an amplitude matrix of this new, bigger size, or to crop the amplitude matrix back to its original size (Fig.\ \ref{beam-rotation-figure}).
The components that use beam rotation in order to simulate the effect of rotated components internally work with the bigger amplitude matrix and then crop the amplitude matrix back to its original size.


\subsection{Beam splitter}

\noindent
The \texttt{Beam splitter} component is an idealisation of a beam-splitter cube with two inputs and two outputs.
If the amplitude-matrix element with indices $i$ and $j$ of input 1 is $u_{\mathrm{in},1}(x_i, y_j)$ and that of input 2 is $u_{\mathrm{in},2}(x_i, y_j)$, then the amplitude-matrix element $(i, j)$ of outputs 1 and 2 is respectively given by
\begin{equation}
u_{\mathrm{out},1}(x_i, y_j) = \frac{u_{\mathrm{in},1}(x_i, y_j) + u_{\mathrm{in},2}(x_i, y_j)}{\sqrt{2}}
\end{equation}
and
\begin{equation}
u_{\mathrm{out},2}(x_i, y_j) = \frac{u_{\mathrm{in},1}(x_i, y_j) - u_{\mathrm{in},2}(x_i, y_j)}{\sqrt{2}}.
\end{equation}

\subsection{Clone of component}

\noindent
The \texttt{Clone of component} is simply precisely the same as another component somewhere in the optical system.

\subsection{Cylindrical lens}

\begin{figure}
\begin{center}
\includegraphics{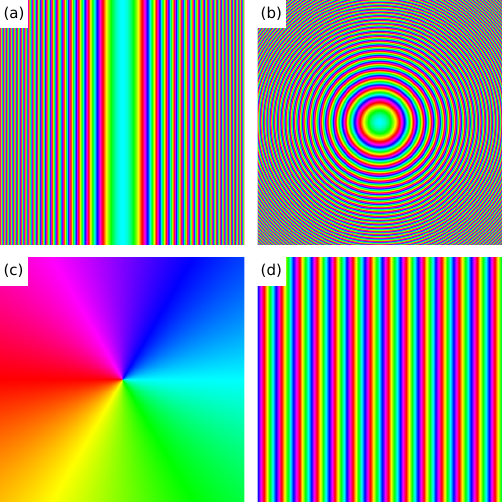}
\end{center}
\caption{\label{phase-elements-figure}Colour representation of the spatial distribution of the phase shift introduced by
(a)~a cylindrical lens, $f = 1\,\mathrm{m}$, cylinder axis at $10^\circ$ to the vertical direction;
(b)~a lens, $f = 1\,\mathrm{m}$;
(c)~a spiral phase plate, $m = 1$;
(d)~a wedge, deflection angle $(x, z)$ projection $0.1^\circ$; and
(e)~a lenslet array with array period $2\,\mathrm{mm}$ and focal length $f = 0.5\,\mathrm{m}$, with its centre (more precisely, the principal point of one of its lenses) located at $(x,y) = (0.5\,\mathrm{mm}, 1\,\mathrm{mm})$.
}
\end{figure}

\noindent
The \texttt{Cylindrical lens} --- like the other glass elements represented in \emph{Young TIM}, \texttt{Lens}, \texttt{Spiral phase plate}, and \texttt{Wedge} --- shifts the phase of each element by an amount proportional to the thickness of the glass at the position corresponding to the element.

For a cylindrical lens of focal length $f$, the phase change is approximately \cite{Goodman-1996-lens-phase}
\begin{equation}
\Delta \Phi = -\frac{k r^2}{2 f},
\end{equation}
where $k = 2 \pi / \lambda$ is the wavenumber of the light and $r$ is the distance from the lens axis.
Fig.\ \ref{phase-elements-figure}(a) shows the spatial distribution of one example of this phase change.

\subsection{Cylindrical-lens mode converter}

\noindent
The \texttt{Cylindrical-lens mode converter} is an arrangement of cylindrical lenses which converts Hermite-Gaussian beams into Laguerre-Gaussian beams and vice versa \cite{Beijersbergen-et-al-1993}.
It takes two arguments: the beam-waist size $w_0$ it works for, and the angle between the cylinder axes of the cylindrical lenses and the $x$ axis.

Specifically, the \texttt{Cylindrical-lens mode converter} represents a $\pi/2$ converter as described in Ref.\ \cite{Beijersbergen-et-al-1993} (which also describes a $\pi$ converter, which, in idealised form, is equivalent to a Dove prism).
A $\pi/2$ converter consists of two parallel cylindrical lenses sharing the same optical axis, each of focal length $f$, separated by a distance $f \sqrt{2}$.
If a Hermite-Gaussian beam with mode indices $m$ and $n$, with waist size
\begin{equation}
w_0 = \sqrt{\left( 1 + 1/\sqrt{2} \right) f \lambda / \pi},
\end{equation}
with its waist plane positioned half-way between the two cylindrical lenses,
and with the nodes orientated with the horizontal and vertical directions
is incident on a $\pi/2$ converter in which the direction of the cylinder axes is at $45^\circ$ with respect to the $x$ axis, then the beam leaving the converter is a Laguerre-Gaussian mode with the same waist size and waist-plane position, and with mode indices
\begin{equation}
l = n - m, \quad
p = \mathrm{min}(n, m).
\end{equation}

A \texttt{Cylindrical-lens mode converter} can be converted into the individual \texttt{Cylindrical lens}es by right-clicking on the component logo and selecting \texttt{Convert to series of optical components} from the pop-up menu.

\subsection{Cylindrical-lens spiral}
\label{s:cylindricalLensSpiral}

\noindent
The \texttt{Cylindrical-lens spiral} is a phase hologram of component based on a cylindrical lens that is bent into a spiral, either a logarithmic spiral, an Archimedean spiral, or a hyperbolic spiral.
In polar coordinates, the radial coordinate $R$ of the cylindrical lens's nodal line (defined as the line that does not deflect light rays incident on it) of the $n$th winding for azimuthal coordinate $\varphi$ is described by the equation
\begin{align}
R(n, \varphi) = \left\{ 
\begin{array}{ccc}
 \exp(b \theta) & :  & \mbox{logarithmic spiral}  \\
b \theta  & : & \mbox{Archimedean spiral} \\
-1/(b \theta)  & :  & \mbox{hyperbolic spiral}  
\end{array}
\right.,
\end{align}
where $b$ is a parameter that determines the tightness of the spiral, $\theta = \varphi + n 2 \pi + \theta_0$, and where $\theta_0$ determines the azimuthal orientation.
The focal length of the cylindrical lens changes as a function of $\theta$.
Each winding can also incorporate one half of a radial Alvarez-Lohmann lens \cite{Alvarez-1978,Lohmann-1970}, as well as an additional phase term that depends on $\theta$.
A suitable combination of two \texttt{Cylindrical-lens spiral}s forms a \texttt{Spiral adaptive Fresnel lens} (see Sec.\ \ref{s:spiralAdaptiveFresnelLens}).

\subsection{\label{distance-subsection}Distance}

\noindent
\emph{Young TIM} represents any (non-zero) separation between the planes of neighbouring optical components by a special ``component'' of type \texttt{Distance}.
Its main parameter is the distance separating the components.

By default, \emph{Young TIM} simulates propagation through a distance using a (non-paraxial) version of a standard Fourier algorithm~\cite{Sziklas-Siegman-1975}.
As there is hardly any optical system that does not involve propagation through some distance, and as beam propagation is also quite prone to problems, we discuss this algorithm here in some detail.

Laser beams usually have an approximately uniform polarisation, which allows us to describe them as scalar waves.
They are also usually monochromatic.
If the complex amplitude of a three-dimensional, monochromatic, scalar wave (which is usually a good approximation to a laser beam) is known in one plane, then it is possible to calculate its complex amplitude everywhere else.
To understand how this works, we consider here how to ``propagate'' the wave into a parallel plane.

As we are considering a scalar wave that is monochromatic, the Fourier transform (or plane-wave decomposition) of its complex amplitude consists exclusively of uniform plane waves of the form
\begin{equation}
u_{k_x,k_y,k_z}(x,y,z) = u_0 \exp[\rmi (k_x x + k_y y + k_z z)],
\label{plane-wave-decomposition-1}
\end{equation}
with the monochromaticity condition
\begin{equation}
k_x^2 + k_y^2 + k_z^2 = \left( \frac{2 \pi}{\lambda} \right)^2.
\label{monochromaticity-condition}
\end{equation}

We now look at one of these plane-wave components.
In the plane $z=0$, it is of the form
\begin{equation}
u_{k_x,k_y,k_z}(x,y,0) = u_0 \exp[\rmi (k_x x + k_y y)];
\end{equation}
in a parallel plane, $z = \Delta z$, it is of the form
\begin{equation}
\begin{aligned}
u_{k_x,k_y,k_z}(x,y,\Delta z)
&= u_0 \exp[\rmi (k_x x + k_y y + k_z \Delta z)] \\
&= u_{k_x,k_y,k_z}(x,y,0) \exp(\rmi k_z \Delta z).
\end{aligned}
\end{equation}
In other words, it is very easy to ``propagate'' a plane wave from one plane into another, parallel, plane:  all that is required is multiplication by $\exp(\rmi k_z \Delta z)$, where $\Delta z$ is the separation between the planes.

Now we generalize from plane waves to much more general monochromatic scalar waves.
If we know the complex amplitude of the wave in the plane $z=0$, i.e.\ we know $u(x,y,0)$, we can still make use of the fact that plane waves are easy to propagate, as follows.
\begin{enumerate}

\item We write $u(x,y,0)$ in terms of its plane-wave components:
\begin{equation}
\begin{split}
& u(x,y,0) = \\
& \quad \frac{1}{2 \pi} \iint \rmd k_x \rmd k_y \tilde{u}_0(k_x,k_y) \exp[\rmi (k_x x + k_y y)].
\end{split}
\end{equation}
We can calculate the amplitudes $\tilde{u}_0(k_x,k_y)$ of the plane-wave components by Fourier-transforming $u(x,y,0)$:
\begin{equation}
\begin{split}
& \tilde{u}_0(k_x,k_y) = \\
& \quad \frac{1}{2 \pi} \iint \rmd x \, \rmd y \, u(x,y,0) \exp[-\rmi (k_x x + k_y y)].
\end{split}
\end{equation}

\item The 2D plane-wave component $\tilde{u}_0(k_x,k_y) \exp[\rmi (k_x x + k_y y)]$ has to be the cross-section in the plane $z=0$ of the 3D plane wave $\tilde{u}_0(k_x,k_y) \exp[\rmi (k_x x + k_y y + k_z z)]$, where $k_z$ can be calculated from the monochromaticity condition (Eqn (\ref{monochromaticity-condition})) and the assumption that the beam travels into the positive $z$ direction, i.e.\ that $k_z$ is positive, so that we know which square root to take.
Then
\begin{equation}
k_z(k_x, k_y) = + \sqrt{\left(\frac{2 \pi}{\lambda}\right)^2 - k_x^2 - k_y^2}.
\label{kz-equation}
\end{equation}

\item We can then calculate the amplitude of this plane wave in a later plane $z = \Delta z$ simply by multiplying it with $\exp(\rmi k_z \Delta z)$.
This can be done by calculating a modified plane-wave amplitude
\begin{equation}
\tilde{u}_{\Delta z}(k_x,k_y) = \tilde{u}_0(k_x,k_y) \exp \left(\rmi 
k_z(k_x, k_y) \Delta z \right).
\label{plane-wave-propagation-equation}
\end{equation}

\item By calculating the superposition formed by the amplitudes of all the plane waves in the later plane $z = \Delta z$, i.e.\ by calculating the inverse Fourier transform of $\tilde{u}_{\Delta z}(k_x, k_y)$, we get the field of the beam in the plane $z = \Delta z$:
\begin{equation}
\begin{split}
&u(x,y,\Delta z) = \\ 
& \quad \frac{1}{2 \pi} \iint \rmd k_x \rmd k_y \tilde{u}_{\Delta z}(k_x,k_y) \exp[\rmi (k_x x + k_y y)].
\end{split}
\end{equation}

\end{enumerate}

This is the basis of Sziklas and Siegman's beam-propagation algorithm \cite{Sziklas-Siegman-1975}.
This algorithm lends itself particularly to implementation in the computer:
\begin{enumerate}
\item The amplitude $u(x,y,0)$ across a square area in the $z=0$ plane is represented by a 2D array of complex numbers, each number representing the complex amplitude at one position.

\item The corresponding array of plane-wave amplitudes $\tilde{u}_0(k_x,k_y)$ can be calculated by a Fast Fourier Transform (FFT; more generally, a discrete Fourier transform, DFT).

\item The elements of that array are individually multiplied by $\exp(\rmi k_z \Delta z)$, resulting in an array representing the plane-wave amplitudes $\tilde{u}_{\Delta z}(k_x,k_y)$ in the plane $z = \Delta z$.

\item An inverse Fast Fourier Transform (IFFT) yields an array representing the amplitudes $u(x,y,\Delta z)$ in the plane $z = \Delta z$.
\end{enumerate}

Note that the term under the square root in Eqn (\ref{kz-equation}) can become negative.
In this case, $k_z$ would become purely imaginary; such a wave is an evanescent wave.
For such waves, the argument of the exponential in Eqn (\ref{plane-wave-propagation-equation}) is a real number, and propagation by a positive distance therefore leads to exponential decay (and propagation through a negative distance to exponential growth, and therefore almost always numerical problems).
The distance over which the amplitude decays to practically zero is usually a few wavelengths.

To avoid the problem of the beam exiting one side of the simulation area and re-entering from the other, it is desirable to stop the beam from reaching the boundary in the first place.
This can be achieved by absorbing, in closely spaced planes, all light close to the boundaries.
This is equivalent to passing the beam through a series of apertures that are completely opaque close to the boundaries and completely transparent everywhere else.
Such boundaries would lead to significant diffractive scattering of the beam.
To avoid this scattering, we use apertures with an absorption profile that varies smoothly 0 to 1 (Fig.\ \ref{f:softBoundaryTransmissionCoefficient}).
This is a variant of the beam-propagation method (BPM) \cite{Van-Roey-et-al-1981}.
If the BPM is selected, propagation proceeds by alternating propagation by a step size and application of the supergaussian absorbing boundary, until the cumulative propagation distance equals the desired overall propagation distance (if necessary, the final propagation step can be less than the step size).
Additional parameters of the BPM are the step size and the width of the absorbing boundary.

\subsection{Dove prism}

\begin{figure}
\begin{center}
\includegraphics{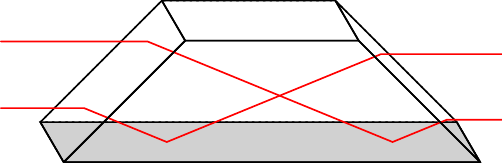}
\end{center}
\caption{\label{Dove-prism-figure}Schematic of two light rays (red lines) passing through a Dove prism.
Each light ray undergoes refraction upon entering the prism, total internal reflection (TIR) at the prism's bottom surface (shaded), and refraction again upon leaving the prism.
In the example shown, one ray is above the other ray upon entering, and below the other ray on leaving.}
\end{figure}

\noindent
Fig.\ \ref{Dove-prism-figure} shows a schematic diagram of a Dove prism, which is essentially a standard $90^\circ$ prism with its top missing.
Passing light through the prism as shown in Fig.\ \ref{Dove-prism-figure} produces a mirror image of the beam (height is inverted).

\emph{Young TIM}'s \texttt{Dove prism} is an idealisation of such a Dove prism.
It simply mirrors the amplitude cross-section with respect to a mirror axis at a given angle with respect to the $x$ axis.
Note that such a component is also an idealised $\pi$ converter (see \texttt{Cylindrical-lens mode converter}) \cite{Beijersbergen-et-al-1993}.

\subsection{Dove-prism array}

\begin{figure}
\begin{center} \includegraphics{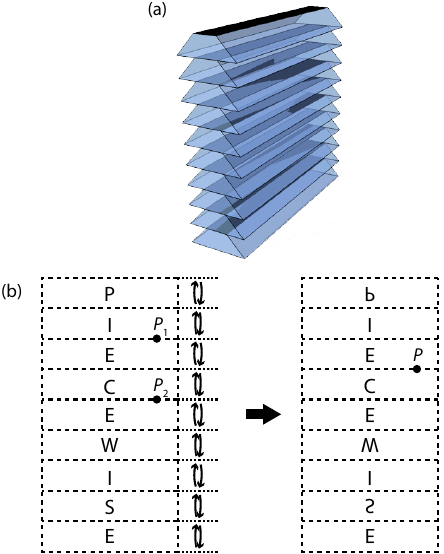} \end{center}
\caption{\label{piecewise-flipping-figure}Dove-prism array and its idealised effect on a transmitted wave front.
(a)~Structure of a Dove-prism array.
(b)~Wave fronts are flipped in strips.
Two parts of the original beam cross-section are marked, $P_1$ and $P_2$, positioned respectively at the top of the third piece and the bottom of the fourth piece (the pieces are counted from the top).
After piecewise flipping, the two parts lie respectively at the bottom of the third piece and the top of the fourth piece, that is, on either side of the boundary between the third and the fourth piece.}
\end{figure}


\noindent
A Dove-prism array\cite{Watkins-2000} 
is a stack of elongated Dove prisms (Fig.\ \ref{Dove-prism-figure}).

\emph{Young TIM}'s \texttt{Dove-prism array} uses the same idealisation as its \texttt{Dove prism}: the piece of the amplitude matrix that enters a particular Dove prism is mirrored with respect to that Dove prism's mirror axis, which is centred on the piece (Fig.\ \ref{piecewise-flipping-figure}; in the \texttt{Dove prism} that piece of the amplitude matrix is the entire amplitude matrix).

\subsection{Fourier lens}

\noindent
A Fourier lens is a standard lens that is used to Fourier-transform the complex amplitude cross-section.
Specifically, the amplitude cross-section in the lens's back focal plane is the Fourier transform of the amplitude cross section in the front focal plane.
The focal length of the Fourier lens matters as it determines the size of the Fourier transform (doubling the focal length doubles the size in each dimension of the Fourier transformed beam cross-section).

A Fourier lens can therefore be represented by propagating the field through a combination of a \texttt{Distance} component with propagation distance $f$, a \texttt{Lens} component with focal length $f$, and another \texttt{Distance} component with propagation distance $f$.
This might be suited to many applications, but all three components are prone to potential problems.

The \texttt{Fourier lens} component models the effect of a Fourier transform differently.
As the desired output is the Fourier transform of the input amplitude cross-section, the \texttt{Fourier lens} simply sets the output field to be the numerically calculated Fourier transform of the input field, and changes the width and height the new beam cross-section represents, taking into account the focal length.
The numerical calculation of the Fourier transform is a combination of a discrete Fourier transform, quadrant swapping (to ensure that each quadrant of the cross-section is in the correct place), and shifting by half an element (using the Fourier shift theorem) to ensure that the centre of the Fourier-transformed beam is located between the four central elements (see Fig.\ \ref{grid-of-points-figure}).

\subsection{Hologram and Hologram (from bitmap)}

\noindent
The \texttt{Hologram} component simply multiplies two amplitude matrices (the two inputs) together, element by element.
If one is interpreted as the beam and the other as spatially-varying multiplication factor for the beam's local amplitude, then this represents the effect of a thin hologram.
Note that these factors are complex numbers, which means the hologram can be a phase hologram, in which case the modulus of all factors is 1;
or an intensity hologram, in which case the argument of all factors is 1;
or a hologram that affects both phase and intensity.

%
Instead of taking two inputs and interpreting one of these as the incident beam and the other as the hologram (like the \texttt{Hologram} component does), the \texttt{Hologram (from bitmap)} component takes only one input, which it interprets as the incident beam, and loads the hologram from a bitmap file.

This bitmap file has to be in BMP format have the same pixel dimensions as the amplitude matrix has elements.
If the brightness of pixel $(i, j)$ is $b_{i, j}$ and its hue is $h_{i, j}$, then the corresponding hologram element is $b_{i, j} \exp[2 \pi (h_{i, j} - 0.5)]$.

\subsection{Hologrammifier}

\noindent
The \texttt{Hologrammifier} produces a hologram from a light beam.
The main parameter is the hologram type, and it can take on the following values:
\begin{description}

\item[Phase]
With this setting, the \texttt{Hologrammifier} produces a phase hologram from the input beam.
The elements of the phase hologram are
\begin{equation}
f_{i, j} = \exp[\rmi p \arg(u_{i, j})],
\end{equation}
where $p$ is an arbitrary real factor that scales the phase step-height and $u_{i, j}$ is element $(i, j)$ of the complex amplitude cross-section of the input beam.
This allows modelling of phase-height mismatch of phase holograms or the use of computer-controlled phase holograms (spatial light modulators, SLMs).

\item[Intensity]
If this setting is selected, the \texttt{Hologrammifier} creates a simple intensity hologram from the input beam.
Element $(i, j)$ of this hologram is simply
\begin{equation}
f_{i, j} = |u_{i, j}|^2,
\end{equation}
i.e.\ the local intensity of the beam.
This represents the way in which intensity holograms were first produced, namely by exposing holographic film with a light beam, and then developing it into a hologram.

\end{description}


\subsection{Lens}

\noindent
The \texttt{Lens} simply introduces a phase change that is proportional to the local thickness of the glass.
Using the standard quadratic approximation to the (usually spherical) lens shape, the phase change introduced by a lens of focal length $f$ at distance $r$ from the centre is \cite{Goodman-1996-lens-phase}
\begin{equation}
\Delta \Phi = -\frac{k r^2}{2 f}.
\label{e:lensPhase}
\end{equation}
Fig.\ \ref{phase-elements-figure}(b) shows a plot of the phase shift introduced by a specific lens.

\subsection{Lenslet array}

\noindent
The \texttt{Lenslet array} simulates transmission through the phase hologram of a square array of identical lenses.
Under the bonnet, the component identifies, for each pixel, first which lens in the square array it corresponds to, then calculates the distance $r$ of the pixel from the centre of that lens, and finally it phase shifts the field at that pixel by $\Delta \Phi$ as given in Eqn (\ref{e:lensPhase}).

Fig.\ \ref{phase-elements-figure}(e) shows a plot of the phase shift introduced by a specific lenslet array.

\subsection{Mirror}

\begin{figure}
\begin{center} \includegraphics{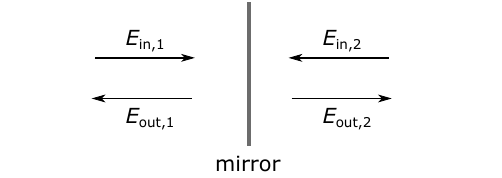} \end{center}
\caption{\label{f:mirror}Schematic of the fields incident on the two sides of a mirror, and the fields leaving the mirror from its two sides.}
\end{figure}

\noindent
The \texttt{Mirror} simulates the effect of two fields being incident on the two sides of a mirror (Fig.\ \ref{f:mirror}).
If the amplitude-matrix element with indices $(i, j)$ of the incident fields are $u_{\mathrm{in}, 1}(x_i, y_j)$ (incident from side 1) and $u_{\mathrm{in}, 2}(x_i, y_j)$ (incident from side 2), then the corresponding amplitude-matrix elements in the outgoing fields $u_{\mathrm{out},1}(x_i, y_j)$ (leaving side 1) and $u_{\mathrm{out},2}(x_i, y_j)$ (leaving side 2) are given by the equations (see Fig.\ 5 in Ref.\ \cite{Bond-et-al-2016})
\begin{align}
u_{\mathrm{out},1}(x_i, y_j) &= r u_{\mathrm{in},1}(x_i, y_j) + \rmi t u_{\mathrm{in},2}(x_i, y_j), \\
u_{\mathrm{out},2}(x_i, y_j) &= r u_{\mathrm{in},2}(x_i, y_j) + \rmi t u_{\mathrm{in},1}(x_i, y_j),
\end{align}
where $r$ is the (amplitude) reflection coefficient, and $t$ is the (amplitude) transmission coefficient.
If the mirror is lossless, $r$ and $t$ are related through the equation (see Ref.\ \cite{Bond-et-al-2016} below Eqn (2.14))
\begin{align}
r^2 + t^2 = 1.
\end{align}

\subsection{Neutral-density filter}

\noindent
The \texttt{Neutral-density filter} can represent one of a number of more or less realistic components that multiply the intensity in the beam by a uniform factor.

The action of the \texttt{Neutral-density filter} can be one of the following:
\begin{description}

\item[multiply intensity by factor $f$]
If this action is selected, the \texttt{Neutral-density filter} can represent either a neutral-density filter or a (uniform) gain medium.
The parameter $f$ is the factor by which the intensity gets multiplied (the amplitude at each point is multiplied by $\sqrt{f}$); a factor between 0 and 1 represents loss, a factor above 1 represents gain.

\item[optical density $d$]
With this action selected, the \texttt{Neutral-density filter} can again represent either a neutral-density filter or a uniform gain medium, but this time it is characterised in terms of the optical density, $d$.
The amplitude at each point is then multiplied by a factor $10^{-0.5 d}$.

\item[set maximum intensity to $I_\mathrm{max}$]
This action allows the \texttt{Neutral-density filter} to adjust its multiplication factor such that the maximum intensity in the beam is $I_\mathrm{max}$.
This is very useful for applications such as holography. 

\item[set power in beam to $P$]
With this action selected, the \texttt{Neutral-density filter}'s multiplication factor is adjusted such that the overall power in the output beam is $P$.
\end{description}

\subsection{Phase-conjugating surface}

\noindent
The \texttt{Phase-conjugating surface} component simply phase-conjugates each element in the amplitude matrix.

This is an idealisation of real-world phase conjugation setups such as four-wave mixing in non-linear media \cite{Yariv-Pepper-1977}.

\subsection{\label{plane-appendix}Plane}

\noindent
The output of a \texttt{Plane} is the same as its input.
\texttt{Plane} components can be useful for the following reasons:
\begin{enumerate}
\item The beam can be visualised in a number of different ways (intensity, phase, anaglyph, Fourier spectrum; different transverse or longitudinal planes; ...), analysed (power within radius of a given point, self-similarity of the intensity cross section with propagation (which was used to calculate Fig.\ 6 in Ref.\ \cite{Sroor-et-al-2019}; a detailed explanation can be found there) etc.), and/or saved as a bitmap image.
\item A plane can have images (which are of type \texttt{Image of plane}), which have the same output as the plane itself, which therefore provide a link between different branches of the tree of components that represents the optical system, and which therefore allow different topologies to be realised.
\end{enumerate}


Internally, the object that represents a plane keeps a copy of the beam in the plane.
This means that the visualisation and save methods have access to the complete set of data;
it also means that the representation of an optical system that contains a closed loop, for example a ring resonator, automatically keeps a copy of the beam during the previously-calculated round trip, which then seeds the next round trip.

A number of bitmap images of transverse beam cross-sections that were saved from a \texttt{Plane} component can be read in again by a \texttt{Light source} or a \texttt{Hologram (from bitmap)}. 



\subsection{Spiral adaptive Fresnel lens}
\label{s:spiralAdaptiveFresnelLens}

\noindent
A \texttt{Spiral adaptive Fresnel lens} comprises two \texttt{Cylindrical-lens spiral}s (Sec.\ \ref{s:cylindricalLensSpiral}) which, in combination, form an adaptive Fresnel lens whose focussing power is proportional to the relative rotation angle between the two \texttt{Cylindrical-lens spiral}s~\cite{Armstrong-et-al-2025,Locher-et-al-2025,Locher-et-al-2025b}.

Compared to representing the same component using a pair of \texttt{Cylindrical-lens spiral}s, a \texttt{Spiral adaptive Fresnel lens} has several advantages, including the following:
\begin{enumerate}
\item When entering the parameters of the component, some duplication can be avoided.
\item The parameters of the \texttt{Cylindrical-lens spiral}s can be automatically calculated such that the \texttt{Spiral adaptive Fresnel lens} has a user-defined focal length for a user-defined rotation angle between the components.
\item A \texttt{Spiral adaptive Fresnel lens} can be converted into the individual \texttt{Cylindrical-lens spiral}s by right-clicking on the component logo and selecting \texttt{Convert to series of optical components} from the pop-up menu.
\end{enumerate}

\subsection{Spiral phase plate}

\noindent
A \texttt{Spiral phase plate} imparts to the beam a spatially-varying phase shift
\begin{equation}
\Delta \Phi = m \varphi,
\end{equation}
where $\varphi$ is the azimuthal angle (Fig.\ \ref{phase-elements-figure}(c)).
If the incident phase has a uniform phase, the outgoing beam has a phase singularity with a topological charge $m$ at the centre.

This functionality can be realised physically with a piece of glass (or, more generally, transparent material)  whose thickness is proportional to the azimuthal angle $\varphi$ and independent of the distance from the centre \cite{Beijersbergen-et-al-1994}.
Between $\varphi = 0$ and $\varphi = 2 \pi$, there is an abrupt change in thickness.
The height of this thickness step determines the proportionality factor, $m$, between the azimuthal angle $\varphi$ and the phase change, $\Delta \Phi$.

\subsection{Wedge}

\noindent
A \texttt{Wedge} imparts a linearly varying phase shift to the beam (Fig.\ \ref{phase-elements-figure}(d)).
Such a linearly-varying phase shift deflects light beams.
\emph{Young TIM} parametrizes a \texttt{Wedge} in terms of the deflection angles in the $(x, z)$ and $(y, z)$ projections, $\alpha$ and $\beta$, it imparts to a light beam that is initially travelling in the $z$ direction.
The phase shift of the amplitude-matrix element corresponding to position $(x, y)$ is
\begin{equation}
\Delta \Phi = k (x \sin \alpha + y \sin \beta),
\end{equation}
where $k = 2 \pi / \lambda$.

\section{\label{source-code-modification-appendix}Outline of a few source-code modification tasks}

\noindent
This paper is not primarily about modifying the source code of \emph{Young TIM}.
Nevertheless, \emph{Young TIM} is open source, and in an attempt to encourage extension and improvement of the code, we outline here a few simple source-code-modification tasks.
In all cases, it is usually extremely helpful to study first the existing code.

\subsection{Definition of a new beam type}

\noindent
In the \texttt{javawaveoptics.optics.lightsource} package, provide an implementation of the \texttt{AbstractLightSource} class.
Add this to the \texttt{initialiseLightSources} method in the \texttt{LightSource} class in the \texttt{javawaveoptics.optics.component} package.

\subsection{Definition of a new optical component}

\noindent
Most optical components only have one input and one output.
To add a new optical component of this type, add a new implementation of the abstract \texttt{AbstractSimpleOpticalComponent} class to the \texttt{javawaveoptics.optics.component} package, then link this component into the \texttt{OpticalComponentFactory} class in the same package.

Finally, add a PNG image of size $150 \times 150$ pixels to the folder that contains the compiled code of the classes in the \texttt{javawaveoptics.optics.component} package.
(To do this in Eclipse, which is one of the most popular cross-platform Java editors\footnote{We have used Eclipse to write \emph{Young TIM}, and we have provided Eclipse launch configuration scripts for starting \emph{Young TIM} with different default optical environments.}, add the PNG image to the folder that contains the source code of the \texttt{javawaveoptics.optics.component} package, and ``Refresh'' the project.
Eclipse then automatically copies these images into the folder containing the corresponding compiled code.)

\subsection{Definition of a new startup environment}

\noindent
In the \texttt{javawaveoptics.optics.environment} package, provide an implementation of the abstract \texttt{AbstractOpticalEnvironment} class.
Then, in the \texttt{OpticalEnvironmentFactory} class, define a new value in the list of environment ``types'' that corresponds to the new environment, and extend the series of cases in the \texttt{createOpticalEnvironment} method to include the new type value and the code that returns the corresponding optical environment.

The new environment can then be selected as the startup environment by passing to the \emph{Young TIM} Java applet (or Java application) the new value of the environment type as the \texttt{environment} parameter.

It is possible to control which types of optical components can be added to the optical environment.
To limit the optical components to a specific set of components (such as those available in the undergraduate lab), use the \texttt{disableLightSource}, \texttt{disableNonLightSource}, \texttt{enableLightSource} and \texttt{enableNonLightSource} methods to create the desired set of components available in the workbench.
Once a particular type of component is disabled, it will not appear in the drop-down menu used to add new components.
Note that it is possible to make components of disabled types part of the startup optical environment, but once deleted in the interactive version such components cannot be added again interactively.

\subsection{Creation of a new version with limited functionality}

\noindent
Create a (startup) environment as an implementation of the \texttt{AbstractOpticalEnvironment} class in the \texttt{javawaveoptics.optics.environment} package.
Also add a corresponding workbench (e.g.\ \texttt{LimitedWorkbench}) to the \texttt{javawaveoptics.ui.workbench} package.

\section{\label{non-planar-surfaces-appendix}Non-planar surfaces}

\noindent
\emph{Young TIM} contains functionality that allows the field to be represented on --- and propagated between --- non-planar surfaces.
This functionality is not accessible from the interactive version.
The aim of this appendix is simply to alert potential users to the existence of this functionality, and to give a few details aimed at facilitating the use of this functionality by accessing the source code.

\begin{figure}
\begin{center}
\includegraphics{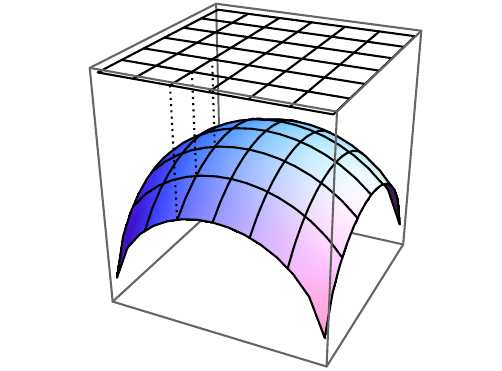}
\end{center}
\caption{\label{non-planar-surface-grid-figure}On a non-planar surface, each value in the amplitude matrix represents the value of the complex field on one of the nodes on a square, planar, grid projected onto the surface.}
\end{figure}

The class representing a beam cross-section on a general surface is \texttt{LightBeamOnSurface2D} in the package \texttt{library.optics}.
It contains an amplitude matrix, the elements of which represent the values of the complex field on a rectangular grid of points projected onto the surface (Fig.\ \ref{non-planar-surface-grid-figure}), in addition to parameters specifying the represented physical size and the wavelength.

So far, the functionality that has been implemented includes propagation to another surface (the ``target surface'') and transmission through a non-planar hologram that images two points into each other.
The former happens in the \texttt{propagate} method, the latter in the \texttt{passThroughImagingHologram} method.
The propagation algorithm places the waist of a small Gaussian beam travelling in the $z$ direction at each grid point; the phase and intensity of the Gaussian beam at the waist are those of the complex field at the corresponding grid point.
These Gaussian beams interfere, and their combined field at the grid points on the target surface provides the new amplitude-matrix values.



\begin{thebibliography}{10}
\newcommand{\enquote}[1]{``#1''}

\bibitem{Huygens-1690}
C.~Huygens, \emph{Trait\'{e} de la Lumi\`{e}re} (Pieter van der Aa, Leiden,
  1690).

\bibitem{Zheludev-Kivshar-2012}
N.~I. Zheludev and Y.~S. Kivshar, \enquote{From metamaterials to metadevices,}
  {{Nat. Mater.}} \textbf{11}, 917--924 (2012).

\bibitem{Ward-Pendry-1996}
A.~J. Ward and J.~B. Pendry, \enquote{Refraction and geometry in {M}axwell's
  equations,} {{J. Mod. Opt.}} \textbf{43}, 773--793
  (1996).

\bibitem{Pendry-et-al-2006}
J.~B. Pendry, D.~Schurig, and D.~R. Smith, \enquote{Controlling electromagnetic
  fields,} {{Science}} \textbf{312}, 1780--1782 (2006).

\bibitem{Leonhardt-2006}
U.~Leonhardt, \enquote{Optical conformal mapping,}
  {{Science}} \textbf{312}, 1777--1780 (2006).

\bibitem{PhET-wave-interference}
{University of Colorado}, \enquote{{PhET} interactive simulations. wave
  interference,}
  \url{https://phet.colorado.edu/sims/html/wave-interference/latest/wave-interference_en.html}
  (accessed 16/12/2021).

\bibitem{Fernandez-et-al-2014}
E.~Fern{\'a}ndez, R.~Fuentes, C.~Garc{\'\i}a, and I.~Pascual,
  \enquote{{Development of Matlab GUI educational software to assist a
  laboratory of physical optics},} in \emph{12th Education and Training in
  Optics and Photonics Conference,}  vol. 9289 M.~F. P. C.~M. Costa and
  M.~Zghal, eds., International Society for Optics and Photonics (SPIE, 2014),
  pp. 484 -- 490.

\bibitem{Baranov-2018}
A.~V. Baranov, \enquote{Students' project developments of wave optics virtual
  labs,} in \emph{2018 XIV International Scientific-Technical Conference on
  Actual Problems of Electronics Instrument Engineering (APEIE),}  (2018), pp.
  240--242.

\bibitem{Yang-et-al-2008a}
B.~Yang, Y.~Huang, R.~Adams, and J.~Z.~K. Burbank, \enquote{Effective teaching
  of photonics e\&m theory using comsol{\textregistered},} in \emph{2008 Annual
  Conference \& Exposition, Pittsburgh, Pennsylvania,}  (2008).

\bibitem{Goodman-1996-lens-phase}
J.~W. Goodman, \emph{Introduction to Fourier Optics} (McGraw-Hill, New York,
  1996), chap. 5.1.3, 2nd ed.

\bibitem{Sziklas-Siegman-1975}
E.~A. Sziklas and A.~E. Siegman, \enquote{Mode calculations in unstable
  resonators with flowing saturable gain. 2: {F}ast {F}ourier transform
  method,} {{Appl. Opt.}} \textbf{14}, 1874--1889 (1975).

\bibitem{Leach-et-al-2002}
J.~Leach, M.~J. Padgett, S.~M. Barnett, S.~Franke-Arnold, and J.~Courtial,
  \enquote{Measuring the orbital angular momentum of a single photon,}
  {{Phys. Rev. Lett.}} \textbf{88}, 257901 (2002).

\bibitem{Hamilton-Courtial-2008a}
A.~C. Hamilton and J.~Courtial, \enquote{Optical properties of a {D}ove-prism
  sheet,} {{J. Opt. A: Pure Appl. Opt.}} \textbf{10},
  125302 (2008).

\bibitem{Hamilton-Courtial-2009}
A.~C. Hamilton and J.~Courtial, \enquote{Metamaterials for light rays: ray
  optics without wave-optical analog in the ray-optics limit,}
  {{New J. Phys.}} \textbf{11}, 013042 (2009).

\bibitem{Courtial-2008a}
J.~Courtial, \enquote{Ray-optical refraction with confocal lenslet arrays,}
  {{New J. Phys.}} \textbf{10}, 083033 (2008).

\bibitem{Bourgenot-et-al-2018}
C.~Bourgenot, E.~Cowie, L.~Young, G.~Love, J.~Girkin, and J.~Courtial,
  \enquote{A new concept of imaging system: telescope windows,}
  {{Proc. {SPIE}}} \textbf{10474}, 104741Z (2018).

\bibitem{YoungTIM-source}
\enquote{Young {TIM} source code,} \url{https://github.com/jkcuk/YoungTIM}
  (2023).

\bibitem{Yim-Lee-1993}
S.~Yim, Donggyu;~Lee, \enquote{Learning wave optics through the computer
  simulation,} {{SNU Journal of Education Research}}
  \textbf{3}, 93--108 (1993).

\bibitem{Lambert-et-al-2012}
D.~Lambert, A.~C. Hamilton, G.~Constable, H.~Snehanshu, S.~Talati, and
  J.~Courtial, \enquote{{TIM}, a ray-tracing program for {METATOY} research and
  its dissemination,} {{Comp. Phys. Commun.}}
  \textbf{183}, 711--732 (2012).

\bibitem{Oxburgh-et-al-2014}
S.~Oxburgh, T.~Tyc, and J.~Courtial, \enquote{{Dr TIM}: Ray-tracer {TIM}, with
  additional specialist scientific capabilities,} {{Comp.
  Phys. Commun.}} \textbf{185}, 1027--1037 (2014).

\bibitem{Wei-et-al-2003}
H.~Wei, X.~Xue, J.~Leach, M.~J. Padgett, S.~M.Barnett, S.~Franke-Arnold,
  E.~Yao, and J.~Courtial, \enquote{Simplified measurement of the orbital
  angular momentum of photons,} {{Opt. Commun.}}
  \textbf{223}, 117--122 (2003).

\bibitem{Beijersbergen-et-al-1993}
M.~W. Beijersbergen, L.~Allen, H.~E. L.~O. van~der Veen, and J.~P. Woerdman,
  \enquote{Astigmatic laser mode converters and transfer of orbital angular
  momentum,} {{Opt. Commun.}} \textbf{96}, 123--132
  (1993).

\bibitem{Padgett-et-al-1996}
M.~J. Padgett, J.~Arlt, N.~B. Simpson, and L.~Allen, \enquote{{An experiment to
  observe the intensity and phase structure of Laguerre-Gaussian laser modes},}
  {{Am. J. Phys.}} \textbf{64}, 77--82 (1996).

\bibitem{Siegman-1986-Fox-Li}
A.~E. Siegman, \emph{Lasers} (University Science Books, Mill Valley,
  California, 1986), pp. 570--571.

\bibitem{Courtial-Padgett-2000b}
J.~Courtial and M.~J. Padgett, \enquote{Monitor-outside-a-monitor effect and
  self-similar fractal structure in the eigenmodes of unstable optical
  resonators,} {{Phys. Rev. Lett.}} \textbf{85},
  5320--5323 (2000).

\bibitem{Watterson-et-al-2003}
C.~M.~G. Watterson, M.~J. Padgett, and J.~Courtial, \enquote{Classic-fractal
  eigenmodes of unstable canonical resonators,} {{Opt.
  Commun.}} \textbf{223}, 17--23 (2003).

\bibitem{Forrester-et-al-2002}
A.~Forrester, M.~L{\"o}nnqvist, M.~J. Padgett, and J.~Courtial, \enquote{Why
  are the eigenmodes of stable laser resonators structurally stable?}
  {{Opt. Lett.}} \textbf{27}, 1869--1871 (2002).

\bibitem{Leith-Upatnieks-1962}
E.~N. Leith and J.~Upatnieks, \enquote{Reconstructed wavefronts and
  communication theory,} {{J. Opt. Soc. Am.}} \textbf{52},
  1123--1130 (1962).

\bibitem{Bryson-2003}
B.~Bryson, \emph{A Short History of Nearly Everything} (Doubleday, 2003),
  chap.~7.

\bibitem{Dodgson-2004}
N.~A. Dodgson, \enquote{Variation and extrema of human interpupillary
  distance,} in \emph{{Stereoscopic Displays and Virtual Reality Systems XI},}
  vol. 5291 A.~J. Woods, J.~O. Merritt, S.~A. Benton, and M.~T. Bolas, eds.
  (2004), pp. 36--46.

\bibitem{Winn-et-al-1994}
B.~Winn, D.~Whitaker, D.~B. Elliott, and N.~J. Phillips, \enquote{Factors
  affecting light adapted pupil size in normal human subjects,}
  {{Investigative Ophthalmology \& Visual Science}}
  \textbf{35}, 1132--1137 (1994).

\bibitem{Berry-Dennis-2000a}
M.~V. Berry and M.~R. Dennis, \enquote{Phase singularities in isotropic random
  waves,} {{Proc. R. Soc. Lond. A}} \textbf{456}, 2059 --
  2079 (2000).

\bibitem{Courtial-et-al-2012}
J.~Courtial, B.~C. Kirkpatrick, and M.~A. Alonso, \enquote{Imaging with complex
  ray-optical refractive-index interfaces between complex object and image
  distances,} {{Opt. Lett.}} \textbf{37}, 701--703 (2012).

\bibitem{Courtial-1999a-beam-propagation-problems}
J.~Courtial, \emph{Angular momentum of light, self-imaging beams, and fractal
  resonator modes} (Ph.D.\ thesis, University of St Andrews, 1999), p. 96ff.

\bibitem{Novak-et-al-1999}
G.~Novak, E.~T. Patterson, A.~D. Gavrin, and W.~Christian, \emph{Just-In-Time
  Teaching: Blending Active Learning with Web Technology} (Prentice Hall, Upper
  Saddle River, NJ, 1999).

\bibitem{Allen-et-al-1992}
L.~Allen, M.~W. Beijersbergen, R.~J.~C. Spreeuw, and J.~P. Woerdman,
  \enquote{Orbital angular momentum of light and the transformation of
  {L}aguerre-{G}aussian modes,} {{Phys. Rev. A}}
  \textbf{45}, 8185--8189 (1992).

\bibitem{Karman-Woerdman-1998}
G.~P. Karman and J.~P. Woerdman, \enquote{Fractal structure of eigenmodes of
  unstable-cavity lasers,} {{Opt. Lett.}} \textbf{23},
  1909--1911 (1998).

\bibitem{Karman-et-al-1999}
G.~P. Karman, G.~S. McDonald, G.~H.~C. New, and J.~P. Woerdman,
  \enquote{Fractal modes in unstable resonators,}
  {{Nature}} \textbf{402}, 138 (1999).

\bibitem{Swift-1972}
D.~W. Swift, \enquote{Image rotation devices -- a comparative survey,}
  {{Opt. Laser Technol.}} \textbf{4}, 175--188 (1972).

\bibitem{Wikipedia-Bilinear-interpolation}
Wikipedia, \enquote{Bilinear interpolation,}
  \url{http://en.wikipedia.org/wiki/Bilinear_interpolation}.

\bibitem{Alvarez-1978}
L.~Alvarez, \enquote{Development of variable-focus lenses and a new refractor,}
  {{Journal of the American Optometric Association}}
  \textbf{49}, 24--29 (1978).

\bibitem{Lohmann-1970}
A.~W. Lohmann, \enquote{A new class of varifocal lenses,}
  {{Appl. Opt.}} \textbf{9}, 1669--1671 (1970).

\bibitem{Van-Roey-et-al-1981}
J.~van Roey, J.~van~der Donk, and P.~E. Lagasse, \enquote{Beam-propagation
  method: analysis and assessment,} {{J. Opt. Soc. Am.}}
  \textbf{71}, 803--810 (1981).

\bibitem{Watkins-2000}
R.~A. Watkins, \enquote{Multiple {D}ove prism assembly,} U. S. Patent
  {6,097,554} (2000).

\bibitem{Bond-et-al-2016}
C.~Bond, D.~Brown, A.~Freise, and K.~A. Strain, \enquote{Interferometer
  techniques for gravitational-wave detection,} {{Living
  Rev. Relativity}} \textbf{19}, 3 (2016).

\bibitem{Yariv-Pepper-1977}
A.~Yariv and D.~M. Pepper, \enquote{Amplified reflection, phase conjugation,
  and oscillation in degenerate four-wave mixing,} {{Opt.
  Lett.}} \textbf{1}, 16--18 (1977).

\bibitem{Sroor-et-al-2019}
H.~Sroor, D.~Naidoo, S.~W. Miller, J.~Nelson, J.~Courtial, and A.~Forbes,
  \enquote{Fractal light from lasers,} {{Phys. Rev. A}}
  \textbf{99}, 013848 (2019).

\bibitem{Armstrong-et-al-2025}
I.~Armstrong, M.~Locher, and J.~Courtial, \enquote{Adaptive {F}resnel lens:
  basic theory,} {{J. Opt. Soc. Am. A}} \textbf{42},
  211--220 (2025).

\bibitem{Locher-et-al-2025}
M.~Locher, D.~Wu, and J.~Courtial, \enquote{Adaptive spiral {F}resnel lens:
  generalisations, improvements, and augmented-reality simulations,} in
  preparation (2025).

\bibitem{Locher-et-al-2025b}
M.~Locher, E.~Blackmore, Z.-K. Wang, and J.~Courtial, \enquote{Wave optics of
  spiral adaptive fresnel lenses,} in preparation (2025).

\bibitem{Beijersbergen-et-al-1994}
M.~W. Beijersbergen, R.~P.~C. Coerwinkel, M.~Kristensen, and J.~P. Woerdman,
  \enquote{Helical-wavefront laser beams produced with a spiral phaseplate,}
  {{Opt. Commun.}} \textbf{112}, 321--327 (1994).

\end{thebibliography}

\end{document}